\documentclass{article}
\usepackage[utf8]{inputenc}
\usepackage[english]{babel}
\usepackage[a4paper, total={6in, 8in}]{geometry}
\usepackage{xcolor}
\usepackage{authblk}
\usepackage{indentfirst}
\usepackage{hyperref}
\usepackage{amsmath}
\usepackage{amssymb}
\usepackage{bm}
\usepackage{bbm}
\usepackage{subcaption}
\usepackage{listings}
\usepackage{verbatim}
\usepackage{color} %red, green, blue, yellow, cyan, magenta, black, white
\definecolor{mygreen}{RGB}{28,172,0} % color values Red, Green, Blue
\definecolor{mylilas}{RGB}{170,55,241}
\usepackage{tikz}
\usetikzlibrary{trees}
\usepackage[nomarkers,figuresonly]{endfloat}

\usepackage{algorithm}
\usepackage{algpseudocode}

\newtheorem{proposition}{Proposition}

\newcommand{\Var}{\mathrm{Var}}

\def\beq{ \begin{equation} }
\def\eeq{ \end{equation} }

\def\square{\vcenter{\vbox{\hrule height .4pt
  \hbox{\vrule width .4pt height 5pt \kern 5pt
        \vrule width .4pt} \hrule height .4pt}}}

\def\sqz{\kern-0.2em}

\newcommand{\E}{\mathbb{E}}

\title{Using birth-death processes to infer tumor subpopulation structure from live-cell imaging drug screening data}
\author[1]{Wu C.}
\author[2]{Gunnarsson E.B.}
\author[3]{Myklebust E.M.}
\author[3,4]{K\"{o}hn-Luque A.}
\author[5,6]{Tadele D.S.}
\author[7,8,9]{Enserink J.M.}
\author[3,4]{Frigessi A.}
\author[2]{Foo J.}
\author[1]{Leder K.}

\affil[1]{%
Department of Industrial and Systems Engineering, University of Minnesota, Twin Cities, MN 55455, USA.}
\affil[2]{%
School of Mathematics, University of Minnesota, Twin Cities, MN 55455, USA.}
\affil[3]{%
Oslo Centre for Biostatistics and Epidemiology, Faculty of Medicine, University of Oslo, 0372 Oslo, Norway}
\affil[4]{%
Oslo Centre for Biostatistics and Epidemiology, Oslo University Hospital, Oslo, Norway}
\affil[5]{%
Department of Medical Genetics, Oslo University Hospital, 0424 Oslo, Norway}
\affil[6]{%
Translational Hematology and Oncology Research, Cleveland Clinic, Cleveland, OH 44131, USA}
\affil[7]{%
Department of Molecular Cell Biology, Institute for Cancer Research, Oslo University Hospital, Oslo, Norway}
\affil[8]{%
Centre for Cancer Cell Reprogramming, Institute of Clinical Medicine, Faculty of Medicine, University of Oslo, Norway}
\affil[9]{%
Section for Biochemistry and Molecular Biology, Faculty of Mathematics and Natural Sciences, University of Oslo, Oslo, Norway}

\date{\today}

\graphicspath{{Plots/}}

\begin{document}
%% MATLAB setting
\lstset{language=Matlab,%
    %basicstyle=\color{red},
    breaklines=true,%
    morekeywords={matlab2tikz},
    keywordstyle=\color{blue},%
    morekeywords=[2]{1}, keywordstyle=[2]{\color{black}},
    identifierstyle=\color{black},%
    stringstyle=\color{mylilas},
    commentstyle=\color{mygreen},%
    showstringspaces=false,%without this there will be a symbol in the places where there is a space
    numbers=left,%
    numberstyle={\tiny \color{black}},% size of the numbers
    numbersep=9pt, % this defines how far the numbers are from the text
    emph=[1]{for,end,break},emphstyle=[1]\color{red}, %some words to emphasise
    %emph=[2]{word1,word2}, emphstyle=[2]{style},    
}

\maketitle

\begin{abstract}
    Tumor heterogeneity is a complex and widely recognized trait that poses significant challenges in developing effective cancer therapies.  In particular, many tumors harbor a variety of subpopulations with distinct therapeutic response characteristics.  
Characterizing this heterogeneity by determining the subpopulation structure within a tumor enables more precise and successful treatment strategies.      In our prior work, we developed PhenoPop, a computational framework for unravelling the drug-response subpopulation structure within a tumor from bulk high-throughput drug screening data. However, the deterministic nature of the underlying models driving PhenoPop restricts the model fit and the information it can extract from the data. As an advancement, we propose a stochastic model based on the linear birth-death process to address this limitation. 
Our model can formulate a dynamic variance along the horizon of the experiment so that the model uses more information from the data to provide a more robust estimation.
In addition, the newly proposed model can be readily adapted to situations where the experimental data exhibits a positive time correlation. We test our model on simulated data (\textit{in silico}) and experimental data (\textit{in vitro}), which supports our argument about its advantages.
\end{abstract}

\section{Introduction}

In recent years the design of personalized anti-cancer therapies has been greatly aided by the use of high throughput drug screens (HTDS) \cite{pauli2017personalized,grandori2018personalized}. In these studies a large panel of drugs is tested against a patient's tumor sample to identify the most effective treatment \cite{pemovska2013individualized,pozdeyev2016integrating,matulis2019functional,bonolo2020direct}. HTDS output observed cell viabilities after initial populations of tumor cells are exposed to each drug at a range of dose concentrations.  The relative ease of performing and analyzing such large sets of simultaneous drug-response assays has been driven by technological advances in culturing patient tumor cells  \textit{in vitro}, and robotics and computer vision improvements.  In principle, this information can be used to guide the choice of therapy and dosage for cancer patients, facilitating more personalized treatment strategies.

However, due to the evolutionary process by which they develop, tumors often harbor many different subpopulations with distinct drug-response characteristics by the time of diagnosis \cite{marusyk2012intra}. This tumor heterogeneity can confound results from HTDS since the combined signal from multiple tumor subpopulations results in a bulk drug sensitivity profile that may not reflect the true drug response characteristics of any individual cell in the tumor.  Small clones of drug-resistant subpopulations may be difficult to detect in a bulk drug response profile, but these clones may be clinically significant and drive tumor recurrence after drug-sensitive populations are depleted. As a result of the complex heterogeneities present in most tumors, care must be taken in the analysis and design of HTDS to ensure that beneficial treatments result from the HTDS.  In recent work  we developed a method, PhenoPop, that leverages HTDS data to probe tumor heterogeneity and population substructure with respect to drug sensitivity \cite{kohn2022phenotypic}. In particular, for each drug, PhenoPop characterizes i) the number of phenotypically distinct subpopulations present, ii)  the relative abundance of those subpopulations and iii)  each subpopulation's drug sensitivity.  This method was validated on both experimental and simulated datasets, and applied to clinical samples from multiple myeloma patients.

\begin{figure}[ht]
    \centering
    \includegraphics[scale = 0.175]{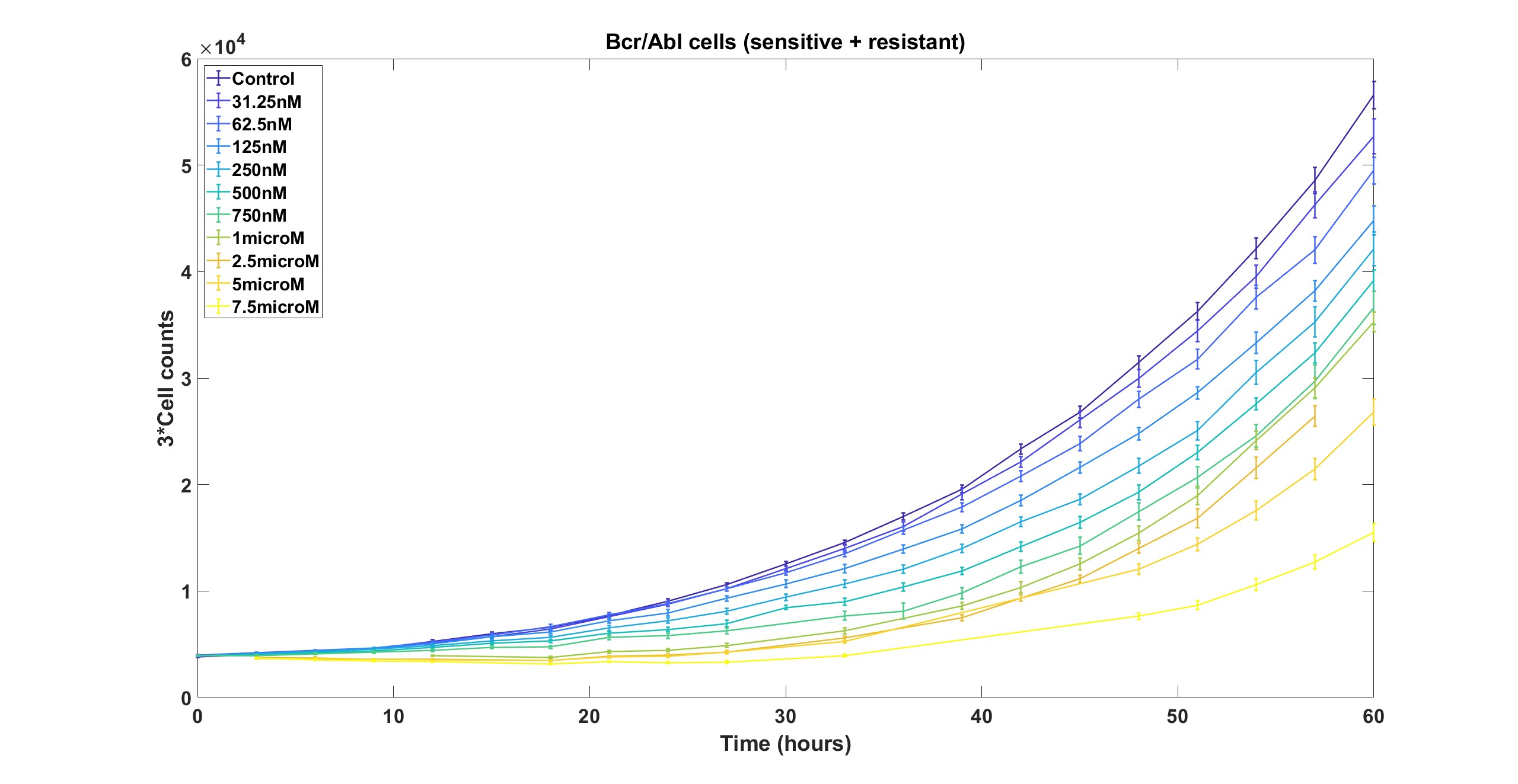}
    \caption{ One to one mixtures of imatinib-sensitive and resistant Ba/F3 cells are counted  at 14 different time points under 11 different concentrations of imatinib. Error bars, based on 14 replicates with outliers removed, depict the sample standard deviation, which increase with larger cell counts.}
    \label{fig:ImatinibData}
\end{figure}

 In the current work, we develop novel theoretical results and computational strategies that  improve PhenoPop by addressing important theoretical and practical limitations.  The original PhenoPop framework was powered by an underlying deterministic population dynamic model of tumor cell growth and response to therapy.   Here we introduce a more sophisticated version of PhenoPop that utilizes stochastic linear birth-death processes, which are widely used to model the dynamics of growing cellular populations  \cite{pakes2003ch,kimmelbranching,Durrett}, as the underlying population dynamic model powering the method. This new framework addresses several important practical limitations of the original approach:  First, our original framework assumed two fixed levels of observational noise; here, the use of an underlying stochastic population dynamic model enables an improved model of observational noise that more accurately captures the characteristics of HTDS data, and reflects the observed dependence of noise amplitude on population size (see Figure \ref{fig:ImatinibData}). Second, this framework allows for natural correlations in observation noise that are tailored to fit specific experimental platforms.  Rather than assuming that all HTDS observations are independent, we may consider data generated using live-cell imaging techniques where the same cellular population is studied at multiple time points, resulting in observational noise that is correlated in time.    By using these stochastic processes to model the underlying populations,  we obtain an improved variance and correlation structure that more accurately models the data and enables more accurate estimators with smaller confidence intervals. 

% {\color{red} It's hard to see where the error bars start and end.} {\color{blue} Put in here a sentence saying that the noise is proportional to the cell count. In general, try to make the figures self-contained, so that someone looking at the figure and not the text knows what the point is.}

The rest of the paper is organized as follows. In Section \ref{sec:Model}, we review the existing PhenoPop method and introduce the new estimation framework based on a stochastic birth-death process model of the underlying population dynamics.  We propose two distinct statistical approaches in the new framework, aimed at analyzing data from endpoint vs. time series (e.g. live-cell imaging) HTDS.  In Section \ref{sec:Experiment}, we conduct a comprehensive investigation of our newly proposed methods and compare them with the PhenoPop method on both \textit{in silico} and \textit{in vitro} data. Finally, we  summarize the results of the investigation and discuss the advantages of the new framework in Section \ref{sec:Discussion}.

\section{Data and model formulation}

\label{sec:Model}
The central problem we address is to infer the presence of subpopulations with different drug sensitivities using data on the drug response of bulk cellular populations.
Here the term `bulk cellular population' refers to the aggregate of all subpopulations within the tumor.
For each given drug, we assume that the data  is in the standard format of total cell counts at a specified collection of time points $\mathcal{T}=\{t_1,\ldots, t_{N_T}\}$ and drug concentrations $\mathcal{D}=\{d_1,\ldots, d_{N_D}\}$.
%for each drug. 
%In particular, assume that the cell counts are observed under drug concentrations $\mathcal{D}=\{d_1,\ldots, d_{N_D}\}$ and the cell count is taken at times in the set $\mathcal{T}=\{t_1,\ldots, t_{N_T}\}$.
Furthermore, assume that for each dose-time pair, $N_R$ 
%statistically identical and 
independent experimental replicates are performed. 
We denote the observed cell count of replicate $r$ at dose $d$ and time $t$ by $x_{t,d,r}$, and denote the total dataset by
$$
\mathbf{x}=\left\{x_{t,d,r}; t\in \mathcal{T}, d\in \mathcal{D}, r\in\{1,\ldots, N_R\}\right\}.
$$

\subsection{PhenoPop for drug response deconvolution in cell populations}

In \cite{kohn2022phenotypic}, we introduced a statistical framework for identifying
the subpopulation structure of a heterogeneous tumor based on drug screen measurements of the total tumor population.
Here, we briefly review the statistical framework and the resulting HTDS deconvolution method (PhenoPop).
First, define the Hill equation with parameters $(b,E,m)$ as 
$$
H(d;b,E,m)=b+\frac{1-b}{1+(d/E)^m},
$$
where $b \in (0,1)$ and $E,m>0$.
A homogeneous cell population treated continuously with drug dose $d$ is assumed to grow at exponential rate $\alpha + \log(H(d;b,E,m))$ per unit time.
If the population has initial size $C_0$, the population size at time $t$ is given by
$$
C_0\exp\left[t\left(\alpha+\log(H(d;b,E,m))\right)\right].
$$
Note that $H(0;b,E,m) = 1$ and $H(d;b,E,m) \to b$ as $d \to \infty$.
Therefore, the population grows at exponential rate $\alpha$ in the absence of drug ($d=0$) and at rate $\alpha+\log(b)<\alpha$ for an arbitrarily large drug dose ($d \to \infty$).
The parameter $E$ represents the dose at which the drug has half the maximum effect, and $m$ represents the steepness of the dose-response curve $d \mapsto H(d;b,E,m)$.

For a heterogeneous cell population, each subpopulation is assumed to follow the aforementioned growth model with subpopulation-specific parameters $\alpha_i$ and $(b_i,E_i,m_i)$.
Assume there are $S$ distinct subpopulations. 
%More precisely, if there are $S$ distinct subpopulations, then 
Then, under drug dose $d$, the number of cells in population $i$ at time $t$ is 
given by 
$$
f_i(t,d)=f_i(0)\exp\left[t\left(\alpha_i+\log(H(d;b_i,E_i,m_i))\right)\right].
$$ 
To ease notation, the dose-response function $H(\cdot;b_i,E_i,m_i)$ for population $i$ will be denoted by $H_i(\cdot)$ in what follows.
%In the above display, $\alpha_i$ is the subpopulation specific cellular growth rate in the absence of the drug and $H(\cdot;b_i,E_i,m_i)$ is the subpopulation specific dose response Hill function, denoted as $H_i(\cdot)$ for simplicity.
The initial size of population $i$ is $f_i(0)=np_i$, where $n$ is the known initial total population size and $p_i$ is the unknown initial fraction of population $i$.
The total population size at time $t$ is then given by
%Then write the total population at time $t$ by
$$
f(t,d)=\sum_{i=1}^Sf_i(t,d)=n\sum_{i=1}^Sp_i\exp\left[t\left(\alpha_i+\log(H_i(d))\right)\right].
$$

A statistical model for the observed data $\mathbf{x}$ is obtained by adding independent Gaussian noise to the deterministic growth model prediction.
%A statistical model for the observed data $\mathbf{x}$ is obtained by adding independent Gaussian noise to the deterministic prediction for each observation.
%To develop a statistical model for the observed data $\mathbf{x}$, we assume independent Gaussian noise with mean zero for each observation. 
%Based on the empirical variance levels in the data, PhenoPop assumes a high variance when the cell population is expected to be above a threshold size (under low drug concentration and when the cells are allowed to grow for a long time) and otherwise uses a lower variance. In particular, define the variance as
The variance of the Gaussian noise is given by
$$
\sigma_{hl}^2(t,d)=\begin{cases}\sigma_H^2,&\enskip t\geq T_L\mbox{ and }d\leq D_L\\
\sigma_L^2,&\enskip\mbox{ otherwise.}\end{cases}
$$
The variance is allowed to depend on time and dose, since at large time points and low doses, a larger variance is expected due to larger cell counts \cite{kohn2022phenotypic}.
Thus, the statistical model for the observation $x_{t,d,r}$ is given by
$$
x_{t,d,r}=f(t,d)+Z^{(r)}(t,d),
$$
where $\{Z^{(r)}(t,d);r\in\{1,\ldots, N_R\}\}$ are independent random variables with the normal distribution $N(0,\sigma_{hl}^2(t,d))$.  
This model has the parameter set
\begin{align}
\label{eq:PP_param_def}
\theta_{PP}(S)=\left\{(p_i,\alpha_i,b_i,E_i,m_i),\sigma_H,\sigma_L;i\in\{1,\ldots,S\}\right\}.
\end{align}
The initial fractions of the $S$ subpopulations $\{p_i:i\in\{1,\ldots S\}\}$ and the parameters $\{(\alpha_i,b_i,E_i,m_i):i\in\{1,\ldots,S\}\}$ governing the drug responses of the subpopulations are unknown. In addition, the variance levels $\sigma_H^2$ and $\sigma_L^2$ are unknown. In practice, the precise values of the thresholds $T_L$ and $D_L$ have minimal effect on the  performance of PhenoPop. 
Therefore, $T_L$ and $D_L$ are treated as known.
%, while the parameters in $\theta_{PP}(S)$ are estimated from the data.
%{\color{red}(WHAT VALUES)}.

The goal of the PhenoPop algorithm is to use the experimental data $\mathbf{x}$ to estimate the unknown parameters $\theta_{PP}(S)$ and the number of subpopulations $S$. 
The parameters $\theta_{PP}(S)$ are estimated via maximum likelihood estimation, where the likelihood function is given by
%PhenoPop employs the maximum likelihood estimation procedure to estimate the parameters $\theta_{PP}(S)$. 
%The likelihood function for the parameters $\theta_{PP}(S)$ given the observations $\mathbf{x}$ is given by
\begin{equation}
\label{eq:PhenoPop Likelihood}
L_{PP}\left(\theta_{PP}(S)|\mathbf{x}\right)=\prod_{r=1}^{N_R}\prod_{(t,d)\in\mathcal{T}\times\mathcal{D}}\frac{1}{\sqrt{2\pi\sigma^2_{hl}(t,d)}}\exp\left[-\frac{\left(x_{t,d,r}-f(t,d)\right)^2}{2\sigma^2_{hl}(t,d)}\right].
\end{equation}
% In \cite{kohn2022phenotypic} we optimized the above likelihood function to derive a maximum likelihood estimator (MLE) for the unknown parameters $\theta_{PP}(S)$. Note that in the above expression, the right hand side depends on the parameter set $\theta_{PP}(S)$ via the function $f(t,d)$.
The likelihood function describes the probability of observing the data $\mathbf{x}$ as a function of the parameter vector $\theta_{PP}(S)$ for a given number $S$ of subpopulations.  The number of subpopulations is then estimated by comparing the negative log likelihood across candidate values of $S$ via the elbow method or Akaike/Bayesian Information criteria. 
For further information, we refer to \cite{kohn2022phenotypic}.

{\it Limitations.} 
The assumption of the PhenoPop algorithm that the Gaussian observation noise has two levels of variance is made for methodological simplicity and does not reflect an observed bifurcation of
%demarcation in 
experimental noise levels. 
It would be more natural to assume that the noise level is directly proportional to the cell count, 
as indicated by the experimental data shown in 
%and in fact, in 
Figure \ref{fig:ImatinibData}.
%it can be seen that the noise levels appear proportional to the cell counts.  
In addition, PhenoPop assumes that all observations are statistically independent. However, if cells are counted using techniques such as live-cell imaging (time-lapse microscopy), then observations of the same well at different time points will be positively correlated.
Both of these limitations can be addressed by modeling the cellular populations with stochastic processes, as we will now show.

\subsection{Linear birth-death process}
A natural extension of PhenoPop \cite{kohn2022phenotypic} is to use a stochatic linear birth-death process to model the cell population dynamics.
In the model, a cell in subpopulation $i$ (type-$i$ cell) divides into two cells at rate $\beta_i \geq 0$ and dies at rate $\nu_i \geq 0$.
This means that during a short time interval of length $\Delta t > 0$, a type-$i$ cell divides with probability $\beta_i \Delta t$ and dies with probability $\nu_i \Delta t$.
%The stochastic model captures fluctuations in cell numbers arising from the stochastic nature of cell division and cell death, and it also imparts a natural time-based correlation structure to the observations, as we discuss in more detail below. 
%As in the PhenoPop method, start with the assumption that there are $S$ distinct subpopulations.  Assume 
%that cells in subpopulation $i$ have birth rate $\beta_i$, and death rate
The death rate of type-$i$ cells is assumed dose-dependent according to
$$
\nu_i(d) = \nu_i - \log(H_i(d))=\nu_i-\log\left(b_i+\frac{1-b_i}{1+(d/E_i)^{n_i}}\right).
$$
The net birth rate $\lambda_i(d) \doteq \beta_i-\nu_i(d)$ of type-$i$ cells is then given by
$$
\lambda_i(d) = (\beta_i-\nu_i) + \log(H_i(d)).
$$
Using the substitution $\alpha_i = \beta_i-\nu_i$, we see that the drug affects the net birth rate of the stochastic model the same way it affects the growth rate $\alpha_i$ of the deterministic population model of PhenoPop.
Note however that here, the drug is assumed to act via a cytotoxic mechanism, that is, higher doses lead to higher death rates. 
Our framework can easily account for cytostatic effects, 
%This assumption can easily be relaxed to also study cytostatic effects 
where higher doses lead to lower cell division rates,
but we focus on cytotoxic therapies for 
%notational 
simplicity.

Let $X_i(t,d)$ denote the number of cells in subpopulation $i$ at time $t$ under drug dose $d$.
The mean and variance of the subpopulation size at time $t$ is given by
%Let $\{X_i(t,d);t\geq 0\}$ be the number of cells in subpopulation $i$, then we can calculate 
\begin{equation}
\label{eq:BD_Mean}
E[X_i(t,d)]\doteq n p_i \mu_i(t,d)=n p_i e^{\lambda_i(d)t}
\end{equation}
\begin{equation}
\label{eq:BD_Variance}
\Var[X_i(t,d)]\doteq n p_i\sigma^2_i(t,d)=n p_i\frac{\beta_i+\nu_i(d)}{\lambda_i(d)}\left(e^{2\lambda_i(d)t}-e^{\lambda_i(d)t}\right).
\end{equation}
%where $\lambda_i(d)=\beta_i-\nu_i(d) \neq 0$ is the net growth rate under $d$ units of drug. Note that if we let $\beta_i - \nu_i = \alpha_i$, it is easy to see that $\lambda_i(d) = \alpha_i + \log(H_i(d))$ is the drug affected growth rate in PhenoPop method. 
Next, denote the total population size at time $t$ under drug dose $d$ by 
$$
X(t,d)=\sum_{i=1}^SX_i(t,d),
$$
with mean and variance
%The mean and variance of the total population size at time $t$ is given by
%For what follows we define the mean and variance of the total population process as
\begin{align*}
E[X(t,d)]\doteq\mu(t,d)&=\sum_{i=1}^Snp_i\mu_i(t,d)\\
\Var[X(t,d)]\doteq n \sigma^2(t,d)&=n\sum_{i=1}^Sp_i\sigma^2_i(t,d).
\end{align*}
Note that the mean size of the total population under the stochastic model equals the total population size under the deterministic model of PhenoPop, again with the substitution $\alpha_i = \beta_i-\nu_i$.
However, the stochastic model introduces variability in the population dynamics at each time point arising from the stochastic nature of cell division and cell death.
%Note that the mean of the stochastic model ahs th
%Note that the mean has the same form as the deterministic growth model in PhenoPop,
%but the 
To account for experimental measurement error, we add independent Gaussian noise to each observation of the stochastic model.
%Note that the mean is the same as in PhenoPop, but the variance is now a more complex function. As in PhenoPop, observational noise corresponding to an independent additive mean zero Gaussian term is incorporated for each observation. 
%However, the variance from the total population process in the linear birth-death model enables us to assume constant variance for observation error at each data point.
As a result, the new statistical model for each observation is 
\begin{align}
\label{eq:stat_model}
x_{t,d,r}=X^{(r)}(t,d)+Z_{t,d,r},
\end{align}
where $X^{(r)}(t,d)$ are independent copies of $X(t,d)$ for $r=1,\ldots,N_R$, and $\{Z_{t,d,r};d\in\mathcal{D},t\in\mathcal{T},r\in\{1,\ldots,N_R\}\}$ are i.i.d.~random variables with the normal distribution $N(0,c^2)$, independent of the $X^{(r)}(t,d)$'s. 
The model parameter set is now
\begin{align}
\label{eq:BD_param_def}
\theta_{BD}(S)=\left\{\left(p_i,\beta_i,\nu_i,b_i,E_i,m_i\right),c;i\in\{1,\ldots,S\}\right\}.
\end{align}
In comparison with PhenoPop, on the one hand, the growth rate parameter $\alpha_i$ for each subpopulation has been replaced by the birth and death rates $\beta_i$ and $\nu_i$.
%there is now an extra parameter per subpopulation.
On the other hand, there is only one parameter $c$ for the observation noise as opposed to four parameters $\{\sigma_H,\sigma_L,T_L,D_L\}$ for PhenoPop.

% P\left(X^{(r)}(t,d)+Z_{t,d,r}=x_{t,d,r}, d\in\mathcal{D},t\in\mathcal{T},r\in\{1,\ldots,N_R\}|\theta_{BD}(S)\right)\\
% &=

Under the new statistical model, the likelihood function is
\begin{align}
\label{eq:BD_LikelihoodFun}
L_{BD}\left(\theta_{BD}(S)|\mathbf{x}\right)&=
\prod_{r=1}^{N_R}\prod_{d\in\mathcal{D}}P\left(X^{(r)}(t,d)+Z_{t,d,r}\in (x_{t,d,r},x_{t,d,r}+\Delta x_{t,d,r}),t\in\mathcal{T}\vert\theta_{BD}(S)\right)
\end{align}
where we assume that observations at different doses and from distinct replicates are independent, and $(x_{t,d,r},x_{t,d,r}+\Delta x_{t,d,r})$ represents an infinitesimally small interval around $x_{t,d,r}$.
We now discuss two different forms this likelihood function can take, depending on whether the data collected at different time points are correlated or not.
%collected at different time points is independent (end-point data) or dependent (live-cell imaging data).

\subsubsection{End-point experiments}
For many common cell counting techniques, e.g.~CellTiter-Glo \cite{hannah2001celltiter}, the experiment must be stopped to perform the viability assay.
In this case, observations at different time points are actually observations of different cell populations exposed to drugs for different amounts of time, and can therefore be treated as independent.
%Thus, there is no correlation between data collected at different time points, and the 
Thus, the likelihood function can be written as
$$
L_{EP}\left(\theta_{BD}(S)|\mathbf{x}\right)=\prod_{r=1}^{N_R}\prod_{d\in\mathcal{D}}\prod_{t\in\mathcal{T}}P\left(X^{(r)}(t,d)+Z_{t,d,r}\in (x_{t,d,r},x_{t,d,r}+\Delta x_{t,d,r})\vert\theta_{BD}(S)\right).
$$
%where the probability on the right-hand side should be viewed as a density.
We note that the distribution of $X^{(r)}(t,d)+Z_{t,d,r}$ can be computed exactly.
%{\color{blue} The right-hand side of the previous display is just the probability density function of a sum of a fixed number of linear birth-death processes and observation noise, and can therefore be evaluated exactly.}
However, for faster computation, one can approximate the distribution by a Gaussian distribution.
%one can replace 
%the distribution with a Gaussian approximation.
%the exact likelihood function of the $X(t,d)$ with a Gaussian approximation. 
To that end, consider the centered and normalized process
\begin{equation}
\label{eq:centered process}
    W_{n}(t,d)=\frac{1}{\sqrt{n}}\sum_{i=1}^S\left(X_i(t,d)-np_ie^{\lambda_i(d)t}\right).
\end{equation}
A straightforward application of the central limit theorem gives the following result.
\begin{proposition}
\label{prop:EndPointCLT}
For $t>0$ and $d\geq 0$, 
$W_{n}(t,d)\Rightarrow N(0,\sigma^2(t,d))$, as $n\to\infty.$
\end{proposition}

Note that `$\Rightarrow$' means converge in distribution. The proof of this result will not be provided since it is a consequence of the more general Proposition \ref{prop:LiveImageCLT}.

Based on Proposition \ref{prop:EndPointCLT},
we obtain the likelihood function
%it is possible to use the likelihood function,
\begin{align}
\label{eq:EndPointsMethod}
L_{EP}\left(\theta_{BD}(S)|\mathbf{x}\right)=\prod_{r=1}^{N_R}\prod_{(t,d)\in\mathcal{T}\times\mathcal{D}}\frac{1}{\sqrt{2\pi(n\sigma^2(t,d)+c^2)}}\exp\left[-\frac{\left(x_{t,d,r}-\mu(t,d)\right)^2}{2\left(n\sigma^2(t,d)+c^2\right)}\right].
\end{align}
The right-hand side of \eqref{eq:EndPointsMethod} depends on the
%growth parameters of the birth-death process 
model parameters $\theta_{BD}(S)$ 
via the mean and variance functions $\mu(t,d)$ and $\sigma^2(t,d)$. As in \cite{kohn2022phenotypic}, one can maximize this expression over the parameter set $\theta_{BD}(S)$ to obtain maximum likelihood estimates of the model parameters. 
%By using $L_{EP}$ instead of $L_{PP}$ there is an increase in the number of model parameters to describe the cell growth dynamics. Note that when using $L_{EP}$ there is a more realistic model for the variance, but it still assumes all observations are independent.
The optimization problem for the new likelihood $L_{EP}$ is more difficult to solve than the corresponding problem for the PhenoPop likelihood $L_{PP}$ in \eqref{eq:PhenoPop Likelihood},
since the variance of the data now depends on the dose-response parameters for the subpopulations.
However, the numerical optimization software we employ is able to deal with this more complex dependence on the model parameters, see Appendix \ref{appx: Environment setting}.

\subsubsection{Live-cell imaging techniques}

Live-cell imaging techniques enable the experimenter to obtain cell counts for the same population across multiple different time points.
%without the need to end the experiment when 
%Live-cell imaging technologies enable viability observations of the same cellular population at multiple time points, since imaging does not require ending the experiment.  
For such datasets, 
observations of the same sample
%observed cell counts 
%from the same sample
%of the same population 
at different time points will be positively correlated. 
In this case, we must compute the joint distribution
%This temporal correlation results in the likelihood: 
\begin{equation}
\label{eq:LC path likelihood}
P\left(X^{(r)}(t,d)+Z_{t,d,r}\in (x_{t,d,r},x_{t,d,r}+\Delta x_{t,d,r}),t\in\mathcal{T}\vert\theta_{BD}(S)\right)
\end{equation}
for each $d\in\mathcal{D}$. 
To ease notation, we will temporarily suppress dependence on the dose.

We first note that $\big(X^{(r)}(t)\big)_{t \geq 0}$ is not a Markov process, since the total cell count at each time point does not include information on the sizes of the individual subpopulations.
Computing \eqref{eq:LC path likelihood} exactly requires summing over the possible sizes of the subpopulations at each time point, which is computationally intensive.
It is possible to speed up the computation
%compute 
%Due to the hidden subpopulations the total population process is not a Markov process and therefore the precise evaluation of the likelihood of a trajectory is a computationally intensive task.
%To evaluate equation 
%\eqref{eq:LC path likelihood} more quickly 
using tools from hidden Markov models, which reduces the computational complexity to $\Omega(\min_{t\in \mathcal{T}} x_{t}^{2})$. 
However, this is still computationally infeasible since $\min_{t\in \mathcal{T}}x_t \approx 5000$, resulting in a 1 second computation time to evaluate a single likelihood.
The computational details 
%behind these observations 
are provided
in Appendix \ref{appx:Exact Path likelihood}.  
%We discuss the details of this approach to computing  \eqref{eq:LC path likelihood}  

A more efficient approach is to use a Gaussian approximation.
For the centered and normalized
%total population 
process $W_{n}(t)$ 
%defined in 
from
\eqref{eq:centered process}, define the vector of observations across time points
%An even more efficien 
%A greater reduction in computational complexity is achieved by introducing a Central Limit Theorem based approximation. In particular, consider the centered and re-scaled total population process $W_{n}(t)$ defined in \eqref{eq:centered process}. Then
%define the observations of the centered process at all the time points by
$$
\mathbf{W}_{n}=\{W_{n}(t);t\in\mathcal{T}\}.
$$
%where for notational simplicity dose dependence has been temporarily suppressed.
By assuming that the set $\mathcal{T}$, number of subpopulations $S$ and initial proportion $p_i$ of each subtype $i$ are independent of the initial total cell count $n$, we derive the following approximation for $\mathbf{W}_{n}$:
\begin{proposition}
\label{prop:LiveImageCLT}
As $n\to\infty$,
$$
\mathbf{W}_{n}\Rightarrow \mathbf{Y}=\{Y(t);t\in\mathcal{T}\}\sim N(0,\Sigma),
$$
where the $(i,j)$ element of the covariance matrix $\Sigma$ is given by
$$
\Sigma_{i,j}=\sum_{\ell=1}^{\min(i,j)}\sum_{k=1}^{S}p_ke^{(\lambda_k t_i-\lambda_kt_\ell)}e^{(\lambda_kt_j- \lambda_kt_\ell)}e^{\lambda_kt_{\ell-1}}\sigma^2_k(t_\ell -t_{\ell-1}).
$$
\end{proposition}

The proof of this proposition is given in Appendix \ref{appx:LiveImageCLT Proof}. In Appendix \ref{appx:LiveImageCLT Extension}, we relax the assumption that the initial proportion $p_i$ is independent of the initial total cell count, and a similar result still follows. In future work, we plan to relax the assumption that $\mathcal{T}$ is independent of $n$.
% Note that in practice, we will define the covariance vector by summing over all sub-populations. This is because in practice we only have $n<\infty$ and we cannot actually assume sub-populations have zero contribution to the variance.

We now reintroduce dose dependence.
%the dose dependence notation and explain how the previous proposition can be used to evaluate our likelihood function. As in the method for end-point data, let $c^2$ denote the variance of our observation noise. 
%In addition, 
For each $d\in\mathcal{D}$, define the $N_T\times 1$ vector 
$$
\mu(d)=\{\mu(t,d);t\in\mathcal{T}\}
$$
and the $N_T\times N_T$ identity matrix $I$. Based on Proposition \ref{prop:LiveImageCLT}, the following approximation is used to compute the likelihood in expression \eqref{eq:BD_LikelihoodFun}:
$$
\mathbf{x}_{\cdot,d,r}=(x_{t,d,r};t\in\mathcal{T})\approx\mu(d)+N(0,n\Sigma(d))+N(0,c^2 I).
$$
The likelihood function is thus given by
\begin{align}
\label{eq:LiveCellLikelihood}
L_{LC}\left(\theta_{BD}(S)|\mathbf{x}\right)=\prod_{r=1}^{N_R}\prod_{d\in\mathcal{D}}\frac{\exp\left[-\frac{1}{2}\left(\mathbf{x}_{\cdot,d,r}-\mu(d)\right)^\top\left(n\Sigma(d)+c^2 I\right)^{-1}\left(\mathbf{x}_{\cdot,d,r}-\mu(d)\right)\right]}{\left(\det\left(2\pi\left(n\Sigma(d)+c^2 I\right)\right)\right)^{1/2}}.
\end{align}
%Importantly note 
Note that the computational complexity of evaluating the above likelihood is independent of $\min_{t\in\mathcal{T}}x_{t}$, alleviating the computational burden associated with an exact evaluation of the likelihood. 
%The issue is that when we attempt to evaluate the likelihood of a trajectory exactly using the Markov property we need to consider all the paths the sub-populations might take that lead to the observed trajectory of the total population. However, when we use the central limit theorem based approach, we no longer rely on the Markov property and thus are no longer concerned with the paths of the sub-populations. Once we know that the entire trajectory is well approximated by a Gaussian vector we can simply write out the approximate likelihood function of the observed trajectory.

The difference between the likelihood function $L_{EP}$ for endpoint data and $L_{LC}$ for live-cell imaging data lies in the structure of the covariance matrix for the observation vector $\mathbf{x}_{\cdot,d,r}$.
%over time.
For $L_{EP}$, observations made at different time points are assumed independent, meaning that the covariance matrix is diagonal.
%Note that the difference between $L_{LC}$ and $L_{EP}$ is in the structure of the covariance matrix. In particular, for $L_{EP}$ the covariance matrix of a cell count trajectory is diagonal, whereas in the live-cell imaging case the covariance matrix is not a diagonal matrix.
%For $L_{LC}$, the time correlation results in a nondiagonal covariance matrix.
For live-cell imaging data, the covariance matrix is not diagonal.
Accurately accounting for time correlations in the likelihood \eqref{eq:LiveCellLikelihood} can improve the accuracy of parameter estimates, as we will discuss in Section \ref{sec:Experiment}.
%will provide some benefits for the accuracy of the parameter inference. However, 
%it is obviously more computationally expensive to calculate the inverses and determinants present in $L_{LC}$. As a result, 
However, it does come at a cost, since it is obviously more computationally expensive to calculate the inverses and determinants present in $L_{LC}$.
As a result, the optimization of $L_{LC}$ can be more difficult than the optimization of $L_{EP}$.

\subsubsection{Accuracy of Gaussian approximation}
%%%%%%%%%%%%%%%%%%%%%%%% We can include the covariance test for different drug dosage %%%%%%%%%%%%%%%%%%%%%%%%%

%{\color{blue} I think it still needs to be clarified what this section adds.}

\begin{comment}
In the derivation of the likelihood function \eqref{eq:LiveCellLikelihood}, the total cell number vector $\mathbf{X}(d)=\{X(t,d);t\in\mathcal{T}\}$ is approximated by a Gaussian distribution $N(\mu(d),n\Sigma(d) + c^2 I)$, based on Proposition \ref{prop:LiveImageCLT}.
In this section, we briefly investigate the quality of this approximation as a function of the initial population size $n$.
\end{comment}

Proposition \ref{prop:LiveImageCLT} states that the centered and normalized process $\mathbf{W}_n$ is approximately Gaussian $N(0,\Sigma(d))$.
However, in the derivation of the likelihood function \eqref{eq:LiveCellLikelihood}, the distribution of the total cell number $\mathbf{X}(d)=\{X(t,d);t\in\mathcal{T}\}$ is approximated with a Gaussian distribution $N(\mu(d),n\Sigma(d) + c^2 I)$, whose mean and variance increases linearly with $n$. 
To verify that the error in this approximation is reasonable for large $n$, we will now compare the distributions of $\mathbf{X}(d)$ and $N(\mu(d),n\Sigma(d) + c^2 I)$ using a well-known measure of the distance between two distributions.
%Due to the normalization term $\sqrt{n}$ in $\mathbf{W}_{n}$, the latter approximation is stronger than the result in Proposition \ref{prop:LiveImageCLT}. It is therefore necessary to numerically compare the distribution of $\mathbf{X}(d)$ and the corresponding Gaussian distribution to see if the approximation is valid. 

%\textbf{Energy distance and its background:}
%%%%%%%%% E-statistic citation need to check %%%%%%%%%%%%%
The energy distance,  introduced in \cite{szekely2003statistics},  is a measure of the distance between probability distributions, which has previously been shown to be related to Cramer's distance \cite{Cramer1928, szekely2003statistics}. The energy distance has been utilized in several statistical tests \cite{BARINGHAUS2004} and is easily computed for multivariate distributions. For probability distributions $F$ and $G$ on $\mathbb{R}^d$, we define their energy distance as
\begin{equation}
\label{eq:Energy}
D(F,G) = \sqrt{2 \E[\|X - Y\|] - \E[\|X - X'\|] - \E[\|Y - Y'\|]},
\end{equation}
where all random variables are independent, $X$ and $X'$, and $Y$ and $Y'$, are distributed according to $F$ and $G$ respectively, and $\|\cdot\|$ denotes the Euclidean norm. 

Since it is unrealistic to compute the equation \eqref{eq:Energy} directly, we approximate the true energy distance by computing the empirical energy distance. For two sets of i.i.d. realization $\{X_1,\cdots, X_k\}, X_i \sim F, \{Y_1,\cdots, Y_m\},Y_i \sim G$, one can obtain the empirical energy distance by
\begin{align}
\label{eq:EmpiricalEnergy}
D_{E}(F,G) = \sqrt{\frac{2}{k m} \sum_{i = 1}^{k} \sum_{j = 1}^{m} \|X_i - Y_j\| - \frac{1}{k^2} \sum_{i = 1}^{k} \sum_{j = 1}^{k} \|X_i - X_j\| - \frac{1}{m^2}\sum_{i = 1}^{m} \sum_{j = 1}^{m} \|Y_i - Y_j\|}.
\end{align}
% {\color{blue} Need some connection between $D$ and $D_E$, or some text explaining why $D_E$ is introduced.}
%As a metric, the empirical energy distance will converge to 0 if and only if $F = G$.

\begin{figure}[ht]
    \centering
    \includegraphics[scale = 0.175]{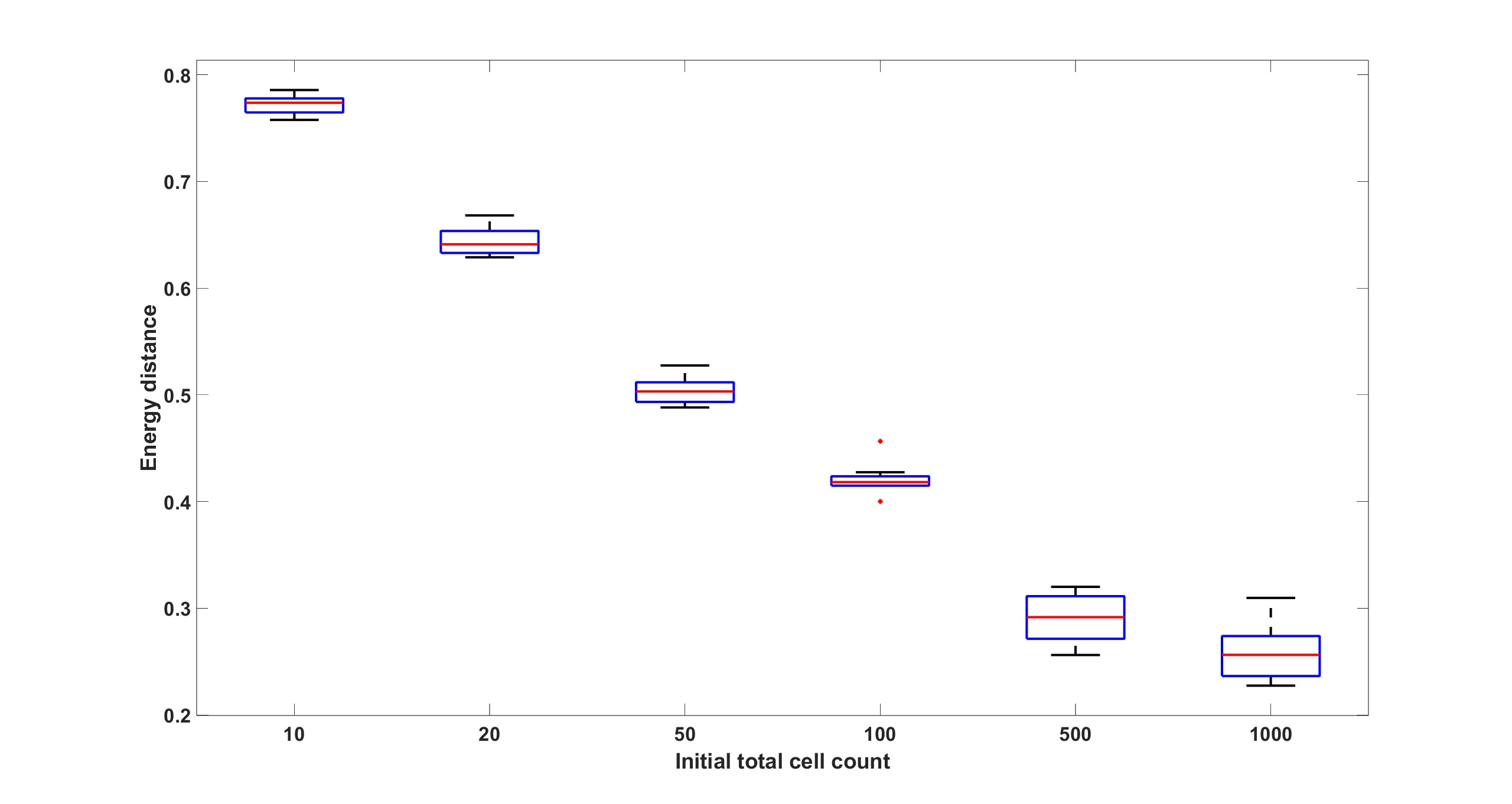}
    \caption{Empirical energy distance between linear birth-death simulated data and multivariate normal distributed data with respect to varying initial cell count: $[10,20,50,100,500,1000]$. The data consists of $N_R=100,000$ replicates and 7 time points $\mathcal{T} = [1,2,3,4,5,6,7]$. No drug effect is assumed. The parameters used to generate the data are $p_1 = 0.4629, \beta_1 = 0.9058,\nu_1 = 0.8101,p_2 = 0.5371, \beta_2 = 0.2785,\nu_2 = 0.2300$. The box plot represents the values from 10 distinct datasets. The figure demonstrates that the distribution of the linear birth-death process converges to the multivariate normal distribution with mean and covariance given by Proposition \ref{prop:LiveImageCLT} as the initial cell count increases.}
    \label{fig:ED}
\end{figure}

%According to the Proposition \ref{prop:LiveImageCLT}, the centered process $\mathbf{W}_n$ will converge to $\mathbf{Y}\sim N(0,\Sigma)$ in distribution as $n\rightarrow \infty$. The next step is to verify that the distribution of $\mathbf{X}(d)$ is well approximated by $N(\mu(d),n\Sigma(d) + c^2 I )$. 
Denote the distribution of $\mathbf{X}(d)$ by $F_{BD}$ and the normal distribution $N(\mu(d),n\Sigma(d) + c^2 I)$ by $F_N$.  Let $\{X_i\}_{i=1}^k$ be $k$ i.i.d.~samples from the distribution $F_{BD}$, and let $\{Y_i\}_{i=1}^m$ be $m$ i.i.d~samples from $F_N$. We can then compute $D_E(F_{BD},F_N)$ using \eqref{eq:EmpiricalEnergy}. In Figure \ref{fig:ED}, we plot $D_E(F_{BD},F_N)$ with varying initial cell counts. The plot shows a monotonic decrease in the empirical energy distance as a function of the initial cell count, which indicates that the distribution of $\mathbf{X}(d)$ is reasonably approximated by a Gaussian distribution for large values of the initial cell count.
%growing closer to a Gaussian distribution as the number of initial cells increases.

\section{Numerical results}
\label{sec:Experiment}

In this section, we use our new statistical methods to analyze both simulated (\textit{in silico}) and experimental (\textit{in vitro}) live cell imaging data.
We apply both the simpler end-point estimation procedure (``end-points method''), based on the likelihood $L_{EP}$ in \eqref{eq:EndPointsMethod}, and the more complex live cell imaging procedure (``live cell image method''), based on the likelihood $L_{LC}$ in \eqref{eq:LiveCellLikelihood}.
%the performance of the inference when using $L_{LC}$ and $L_{EP}$ is investigated.
The performance of the new methods is compared with the existing PhenoPop algorithm.
%(i.e., using the likelihood function $L_{PP}$), utilizing both simulated (\textit{in silico}) and experimental (\textit{in vitro}) data.
In all analyses it is assumed that the observation at time $t=0$ represents the known starting population size, i.e. $x_{0,d,r} = n$.
%When estimating the model parameters using the likelihood function $L_{LC}$, the resulting method is referred to as the "live cell image method." Similarly, when employing $L_{EP}$, the phrase "end-points method" is utilized, and when utilizing $L_{PP}$, the method is referred to as the PhenoPop method.

%%%%%%%%%%%%%%%%%%%%%%%%%%%%%%%%%%%%% 2022/8/24 %%%%%%%%%%%%%%%%%%%%%

%%%%%%%%% Use only one data %%%%%%%%

\subsection{Application to simulated data} 

% \subsubsection{Parameter setting:}
% \label{Parameter setting}
We first apply our estimation methods to simulated ({\em in silico}) data.
In Appendix \ref{appx: Environment setting}, we provide details of the data generation and the parameter estimation for these {\em in silico} experiments.
%computation of maximum likelihood estimates for these
%\textit{in silico} experiments.
%To assess the uncertainty in the estimation, we 
%We perform the maximum likelihood estimation (MLE) on multiple bootstrapped datasets and use that to generate confidence intervals as described (Appendix \ref{appx: Environment setting}).
%a comprehensive description of the data generation and the estimation procedure for the \textit{in silico} experiments.

% {\color{blue} Can this section in blue be moved to Appendix? 

% \noindent 

% %\textbf{Generating parameter:}

% % We select the parameters uniformly from the intervals given in table \ref{Generating_table1}:

% % 

% % Such region for generating parameter was selected to satisfy
% % \begin{itemize}
% %     \item Concentration level $Conc$ cover the $GR_{50}$ of all sub-population cells 
% %     \item The $GR_{50}$ for two distinctive sub-populations do not locate on the same interval of the concentration level $Conc$.
% %     \item Each sub-population has or will have significant amount of population. Otherwise, it may be confused with the noise.
% % \end{itemize}

% \noindent
% }

\subsubsection{Examples with 2 subpopulations} \label{sec:2subpexp}

\label{sec:Illustrative example}
For illustrative purposes, we begin 
%by examining 
with
a case study involving 
an artificial tumor with two subpopulations. 
%In this analysis, we employ estimators based on two different methodologies: the live cell image method with likelihood $L_{LC}$ and the end-points method with likelihood $L_{EP}$. In the present example, we consider a heterogeneous population comprising both drug-sensitive and drug-resistant subpopulations.
%Therefore, we restrict the range of $E_{s},E_{r}$, which control the $GR_{50}$ of each subpopulation, separately  at low and high drug concentration intervals respectively.
Data is generated using a parameter vector $\theta_{BD}(2)$ selected uniformly at random from the ranges in Table \ref{table: Generating_table_2_pop_illustrative}.
We assume that one tumor subpopulation is drug-sensitive and the other is drug-resistant. These subpopulations are indicated by the subscripts $s$ and $r$, respectively.

 \begin{table}[ht]
    \centering
    \begin{tabular}{|c|c|c|c|c|c|c|c|c|c|}
    \hline
         & $p_{s}$&$p_{r}$ & $\beta_{s,r}$ & $\nu_{s,r}$ & $b_{s,r}$ & $E_s$& $E_r$ & $m_{s,r}$& $c$  \\
         \hline
         Range & $[0.3 , 0.5]$ &$1-p_s$& $[0, 1]$ & $[\beta-0.1 , \beta]$ & $[0.8 , 0.9]$ & $[0.05 , 0.1]$  &$[0.75 , 2.5]$& $[1.5 , 5]$ & $[0, 10]$ \\
         \hline
    \end{tabular}
    \caption{Range for parameter generation of experiments with 2 subpopulations}
    \label{table: Generating_table_2_pop_illustrative}
\end{table}

\begin{figure}[ht]
    \makebox[\textwidth][c]{\includegraphics[scale = 0.2]{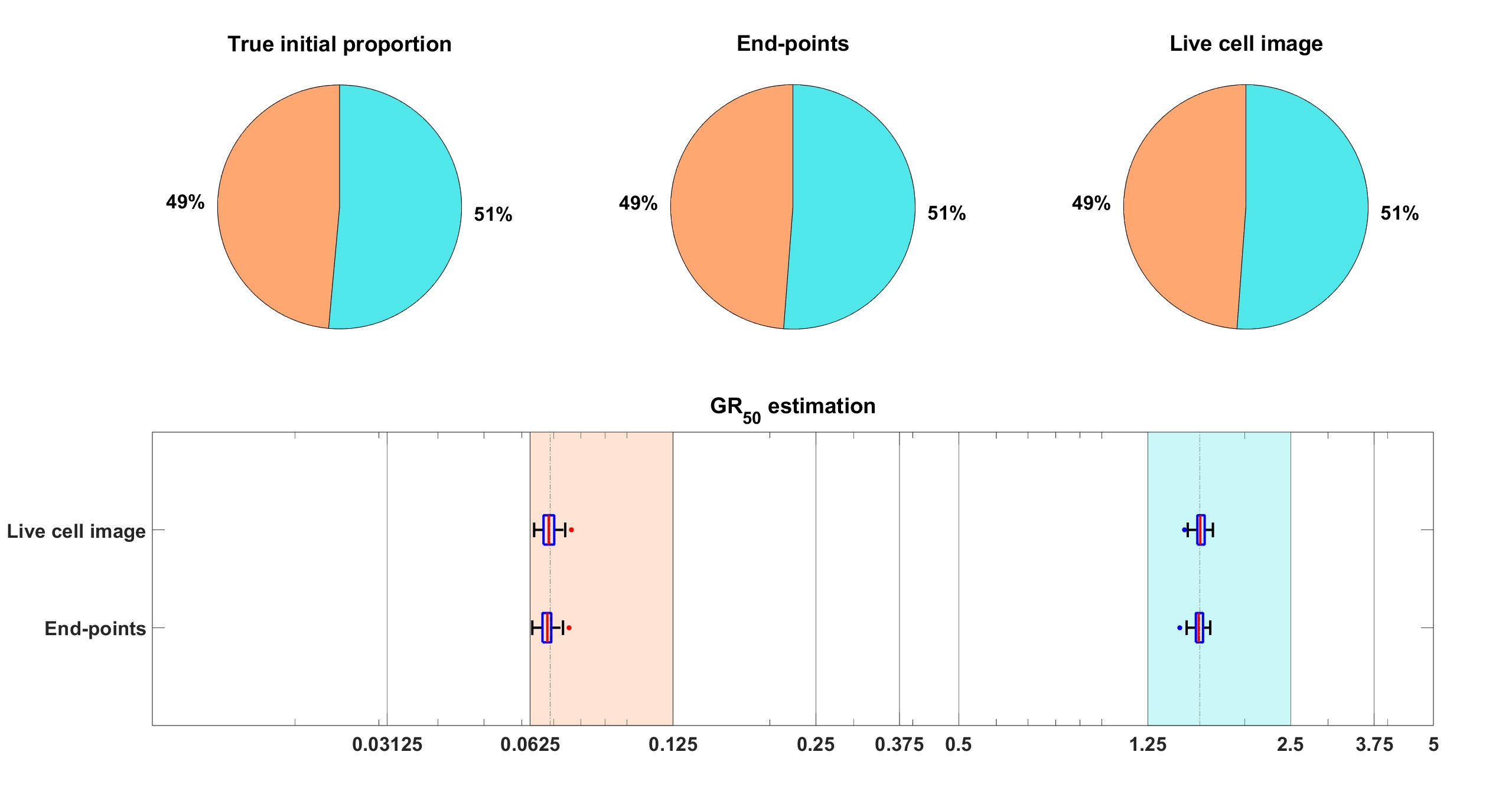}}
    \caption{Estimation of the initial proportion and $GR_{50}$ for 2 subpopulations using the end-points method and the live cell image method on simulated data. The parameter vector $\theta_{BD}(2)$ and observation noise $c$ used in this example are $p_s = 0.4856, \beta_s = 0.1163,\nu_s = 0.0176,b_s = 0.8262,E_s = 0.0674,m_s = 4.5404,p_r = 0.5144,\beta_r = 0.4624,\nu_r = 0.3978,b_r = 0.8062,E_r = 1.5776,m_r = 4.2002,c = 1.2103$. The pie chart illustrates the average of all bootstrap estimates for the initial proportion, while the box plot summarizes the distribution of the estimates for the $GR_{50}$'s. The vertical dashed lines in the box plot correspond to the true $GR_{50}$ values employed to generate the data, while the vertical solid lines indicate the concentration levels at which the data were collected. Each color in the plot represents a distinct subpopulation: orange for sensitive and blue for resistant. The shaded areas in the box plot indicate the concentration intervals where the true $GR_{50}$'s are located, and the colored dots mark outliers in the estimation of the $GR_{50}$ for each subpopulation, with red for sensitive and blue for resistant. This example demonstrates that our newly proposed models can accurately recover the initial proportion and $GR_{50}$ values with high precision.}
    \label{fig:Illustrative}
\end{figure}

As in \cite{kohn2022phenotypic}, 
%for the parameter estimation,
we focus on inferring the initial proportion $p_s$ of sensitive cells,
as well as the $GR_{50}$ dose for each subpopulation.
The $GR_{50}$ is the dose at which the drug has half the maximal effect on the cell death rate,
%dose at which the drug's effect on cell death rate is half the observed maximum,
%the drug has half the maximum effect on cell death rate,
as is further explained in Appendix \ref{appx: Environment setting}.
Informally, the $GR_{50}$ dose for each subpopulation is a measure of the subpopulation's sensitivity to the drug.
To assess the uncertainty in the parameter estimation, we compute maximum likelihood estimates for 100 bootstrapped datasets, as described in Appendix \ref{appx: Environment setting}.
%We perform estimation on 100 bootstrapped datasets, which enables us to assess the uncertainty in the estimation for the $GR_{50}$ doses.
%To check the robustness of the estimator, we perform the maximum likelihood estimation (MLE process on multiple bootstrapped data as described in Appendix \ref{appx: Environment setting} to obtain multiple estimators. Details of parameters are provided in the caption of Figure \ref{fig:Illustrative}.
The results of the case study are shown in Figure \ref{fig:Illustrative},
where we see that both the live cell image method and the end-points method are able to recover the initial proportion $p_s$ of the sensitive population and the $GR_{50}$ dose for each subpopulation accurately.
%, as is shown in Figure \ref{fig:Illustrative}.
%As the figure shows, 
%the results of the parameter estimation using the two newly proposed methods.
%that for the case study data, 
%both the live cell image method and end-points method recover the initial proportion of the sensitive population $p_s$ and the $GR_{50}$ for both subpopulations accurately.
%The figure shows the point estimate for $p_s$, which  is obtained as the average estimate across 100 bootstrapped datasets, while estimates for the $GR_{50}$'s are shown using

%, i.e., the interquartile ranges (IQR) from the box plot cover the true $GR_{50}$ values. 

% {\color{red} what do the light colored vertical bars in the figure mean? the caption is not clear on this.  the caption should not have the word `we' in it, also check for grammar.}
%Next consider a large number of simulated parameter sets to ensure that the methods perform well in a wide range of parameter settings.
We next evaluate the performance of the estimation methods across 30 simulated datasets, where each parameter vector $\theta_{BD}^i(2)$ for $i=1,\ldots,30$ is sampled from the ranges in Table \ref{table: Generating_table_2_pop_illustrative}. 
We furthermore compare the performance of the two new methods with the performance of PhenoPop.
\begin{comment}
We compare the accuracy of estimates for 
%The accuracy of estimation of 
$\{p_s,GR_{s},GR_{r}\}$ using the PhenoPop method ($PP$), end-points method ($EP$), and live cell image method ($LC$).
    For $i \in \{1,\cdots,30\}$ and $j \in \{PP,EP,LC\}$,
%was then computed.
the estimates are denoted by 
\[\hat{\theta}_i^j = \left(\hat{p}_{i,s}^j, \hat{GR}_{i,s}^j, \hat{GR}_{i,r}^j\right).\]
\end{comment}
%For each method and each parameter $\{p_s,GR_{s},GR_{r}\}$, a point estimate is obtained as the mean of maximum likelihood estimates from 100 bootstrapped data sets.
%MLE estimators from 100 Bootstrapping data sets. 
%Given the true parameters $\{p_{i,s},GR_{i,s},GR_{i,r}\}$ from $\theta_{BD}^i(2)$, it is then possible to compute the estimation error. 
The error in the estimation of each parameter $\{p_s,GR_{s},GR_{r}\}$ is measured by considering the absolute log ratio between the point estimate $\hat{x}$ and the true value $x$ for the parameter,
%The absolute log ratio is used to measure the error, i.e.,
\begin{equation}
    \begin{split}
    Er(\hat{x};x) &= \left\lvert\log\left(\frac{x}{\hat{x}} \right) \right\rvert.
    \end{split}
    \label{eq:Log_Ratio}
\end{equation}
\begin{comment}
\begin{equation}
    \begin{split}
    Er_{i}^{j}(p_s) &= \left\lvert\log\left(\frac{p_{i,s}}{\hat{p}_{i,s}^j} \right) \right\rvert\\
    Er_{i}^{j}(GR_{s}) &= \left\lvert\log\left(\frac{GR_{i,s}}{\hat{GR}_{i,s}^j} \right)\right\rvert\\
    Er_{i}^{j}(GR_{r}) &= \left\lvert\log\left(\frac{GR_{i,r}}{\hat{GR}_{i,r}^j} \right)\right\rvert.
    \end{split}
    \label{eq:Log_Ratio}
\end{equation}
\end{comment}
This metric is chosen to address the logarithmic scale associated with the $GR_{50}$ dose.
%These metrics are chosen to address the logarithmic scale of concentration level.

\begin{figure}[ht]
    \centering
    \includegraphics[width = \linewidth]{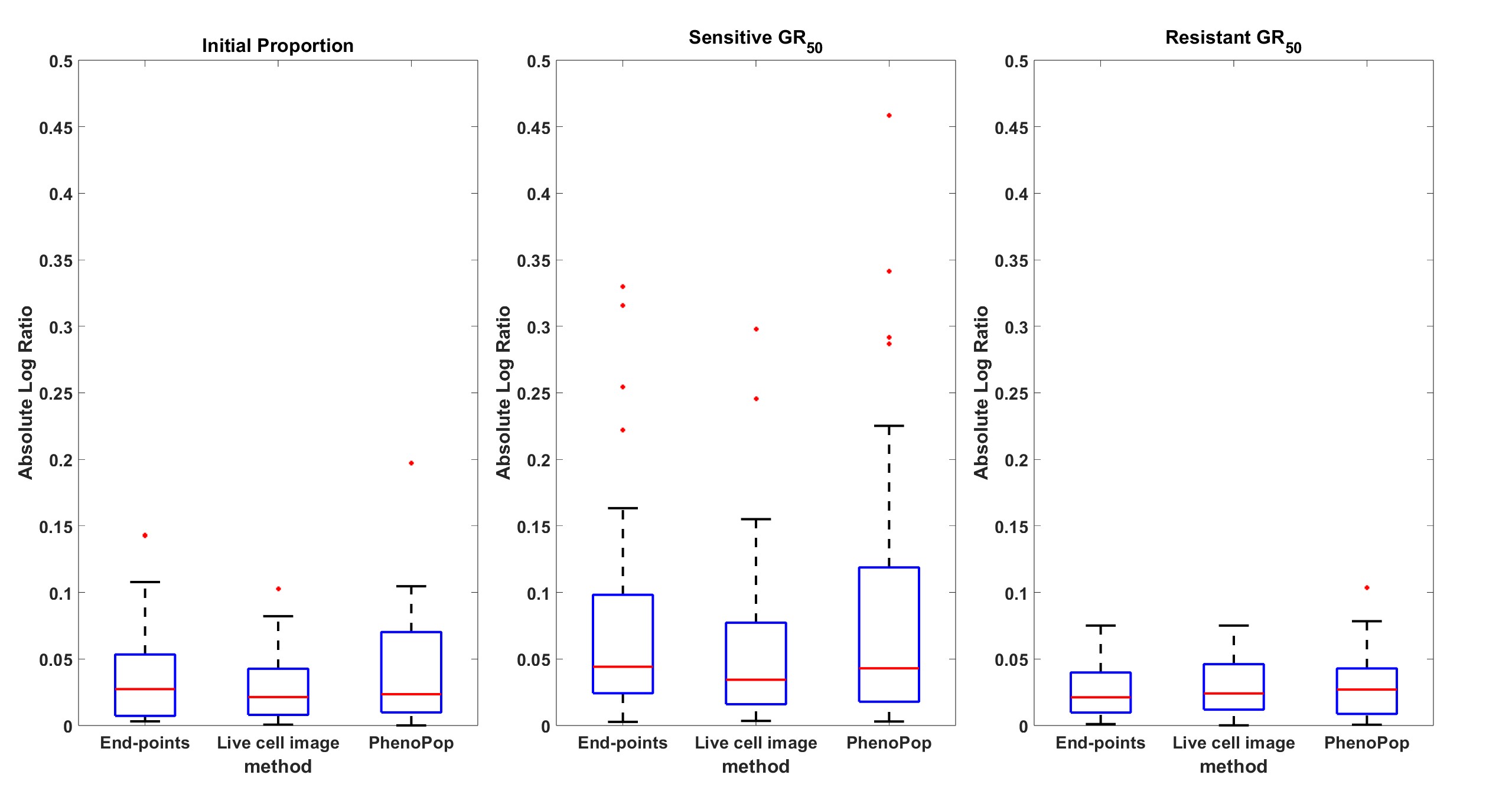}
    \caption{Absolute log ratio accuracy of three estimators $\{\hat{p}_s,\hat{GR}_s,\hat{GR}_r\}$ using the PhenoPop, end-points and live cell image methods. The results are summarized based on 30 different simulated datasets. This figure demonstrates that there are no significant differences in estimation accuracy among these three methods when the true parameters fall within the range described in Table \ref{table: Generating_table_2_pop_illustrative}.}
    \label{fig:Log_Ratio_Accuracy}
\end{figure}

In Figure \ref{fig:Log_Ratio_Accuracy}, a box plot of the estimation errors for the three methods across the 30 datasets is presented.
%a box plot of 
%errors from 30 experiments is presented.
Note that all three parameters 
$\{p_s,GR_{s},GR_{r}\}$
are estimated accurately using all three methods.
%all three methods have good accuracy in estimating all three parameters. 
In addition, the error in estimating the sensitive $GR_{50}$ is larger than the error in estimating the resistant $GR_{50}$ for all three methods. One possible reason is that the initial proportion of sensitive cells is $p_s \in [0.3,0.5]$, so the experimental data contains less information on the sensitive subpopulation. 
Later experimental results will lend further support to this hypothesis.
%will provide evidence that further supports this hypothesis.
%Later experiments will also provide evidence to support this hypothesis. 

% Figure \ref{fig:Log_Ratio_Accuracy} shows that the newly proposed methods do not provide a significant advantage in point estimation accuracy compared to the existing PhenoPop method. 
% However, there is a statistically significant difference in the precision of the estimates, as we will now show.
We next compare the estimation precision of the three methods.
%performance of the live-cell and end points methods with PhenoPop, finding that the new methods improve upon the precision of the parameter estimates.
Specifically, we will compare the widths of the $95\%$ confidence intervals for the three parameters between the three different methods. 
Since $GR_{50}$ values vary significantly across the 30 generated datasets, we normalize the CI width for each method by dividing it by the sum of the CI widths of all three methods for the same dataset.
%within each sample.
%This gives a relative comparison between the CI widths of the three methods across the 30 datasets which is independent of the scale of the parameter in question.

%Due to the random selection of $\theta_{BD}(2)$, $GR_{50}$ values for each subpopulation exhibit variation across the 30 samples.
%Consequently, to account for this variation, we normalize the CI width of each model by dividing it by the sum of CI widths from all three models within each sample. \textcolor{red}{Check if this argument makes sense.}

% For each sample, we normalize its CI width by dividing it by the sum of all three CI widths. \textcolor{red}{I don't understand why we normalize the CI widths. How does that have anything to do with randomness in $\theta_{BD}(2)$???} 

\begin{figure}[ht]
\begin{subfigure}{.33\textwidth}
  \centering
  % include first image
  \includegraphics[width=\linewidth]{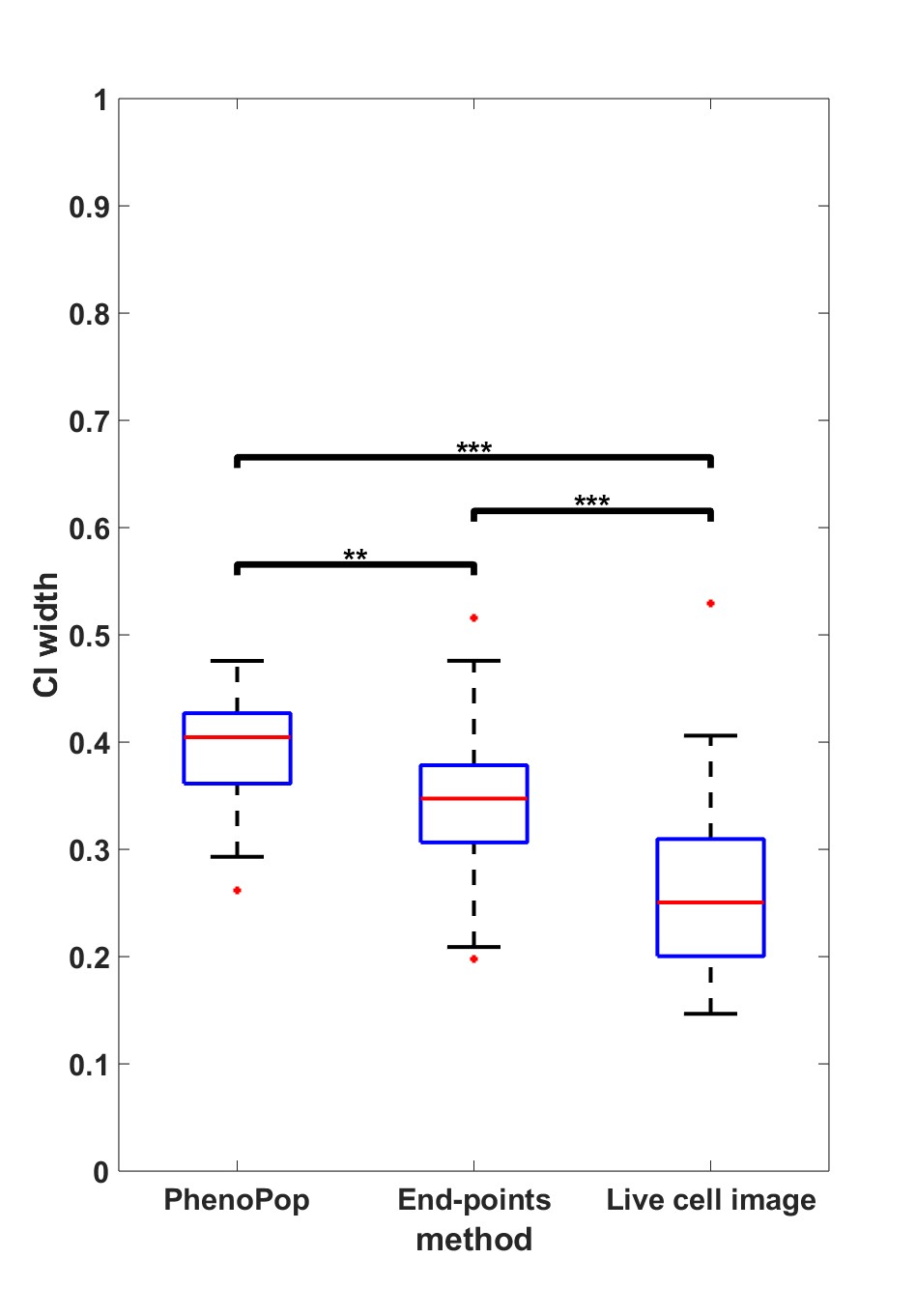}  
  \caption{Initial proportion}
  \label{fig:CI_p}
\end{subfigure}
\begin{subfigure}{.33\textwidth}
  \centering
  % include second image
  \includegraphics[width=\linewidth]{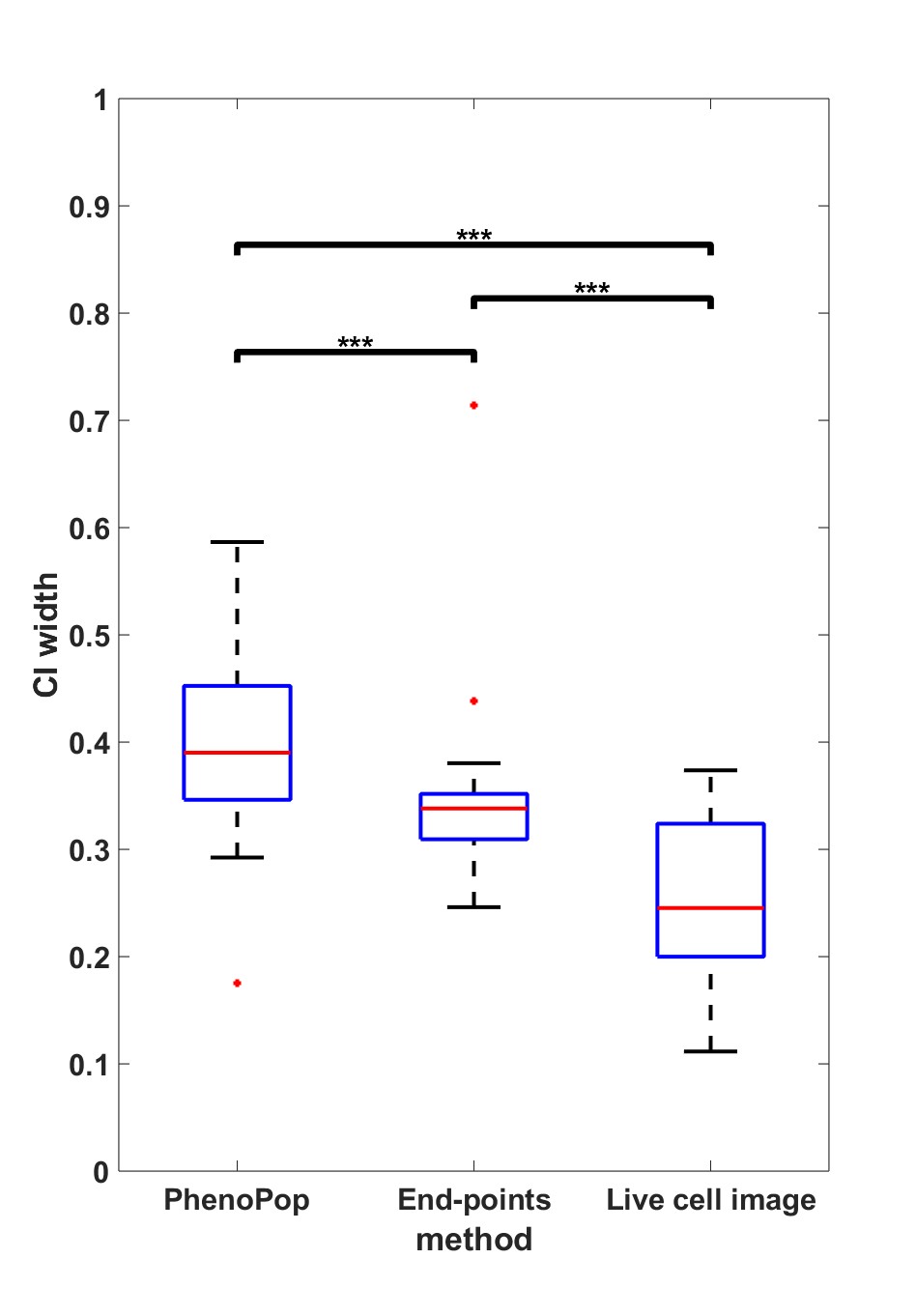}  
  \caption{Sensitive $GR_{50}$}
  \label{fig:CI_GR1}
\end{subfigure}
\begin{subfigure}{.33\textwidth}
  \centering
  \includegraphics[width=\linewidth]{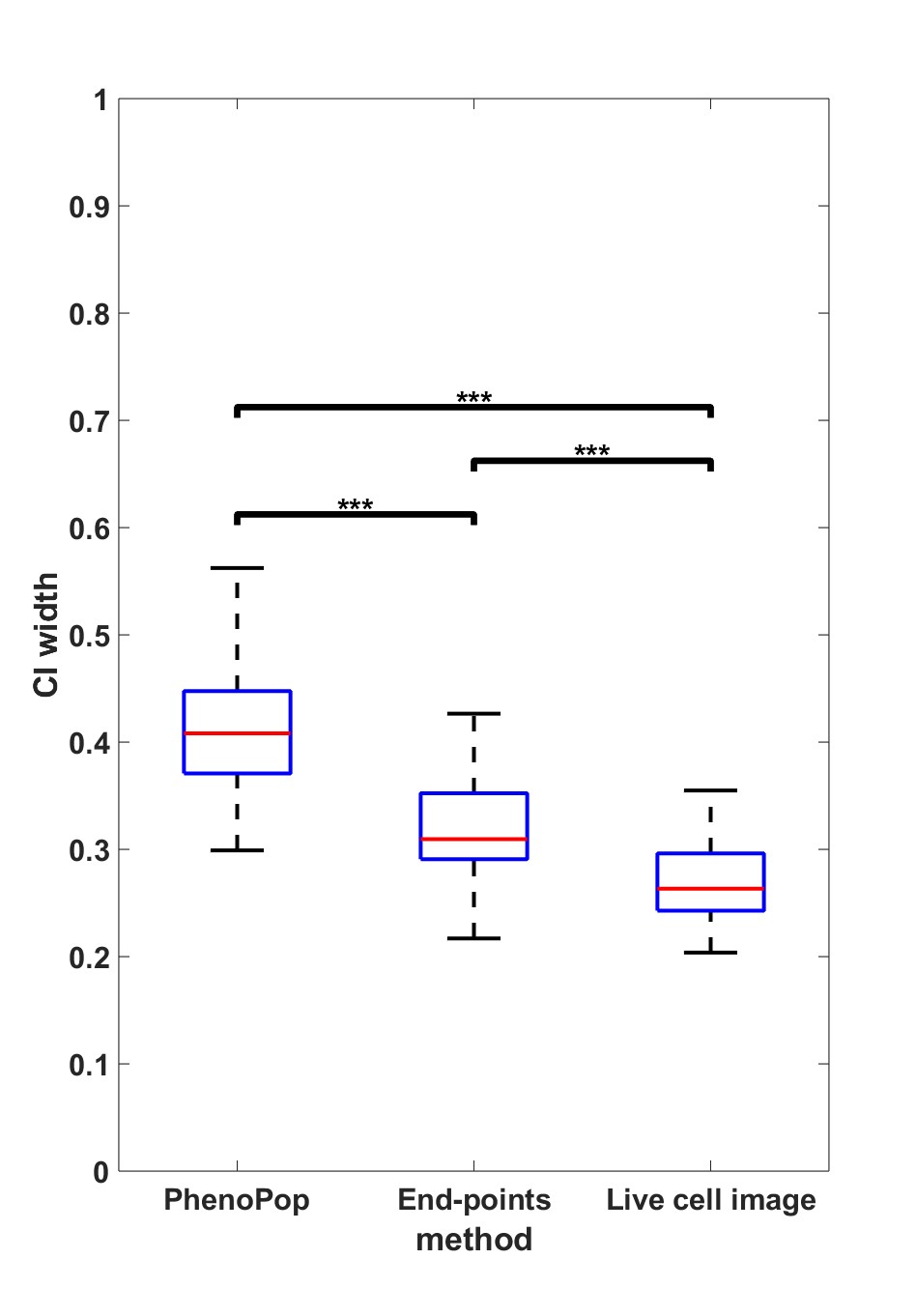}
  \caption{Resistant $GR_{50}$}
  \label{fig:CI_GR2}
\end{subfigure}
\caption{Comparison of the normalized CI widths of three estimators $\{\hat{p}_s, \hat{GR}_s, \hat{GR}_r\}$ estimated from three different methods. The $y$-axis represents the normalized CI width. The box plot summarizes the results across 30 different simulated datasets. The significance bar indicates the p-values derived from the Wilcoxon rank-sum test, with significance levels denoted as $*** \leq  0.001 \leq  ** \leq   0.01 \leq * \leq 0.05$. This figure demonstrates that the newly proposed models exhibit significant advantages in estimation precision, with the live cell image method demonstrating the highest level of precision.}
\label{fig:CI width}
\end{figure}

In Figure \ref{fig:CI width}, the normalized CIs for $\{p_s,GR_{s},GR_{r}\}$ are compared between the three methods across the 30 datasets.
%the normalized confidence interval widths are compared across our three methods for measuring the three quantities: initial proportion, sensitive $GR_{50}$ value, and resistant $GR_{50}$ value. 
First, note that for the initial proportion $p_s$, 
%the initial proportion 
the live cell image method has significantly narrower CIs than the other two methods. Additionally, there is a small but statistically significant difference between the CI widths for the end-points method and the PhenoPop method.
For the sensitive $GR_{50}$ index, the live cell image method again has significantly narrower confidence intervals than the other two methods, and the end-points method has significantly narrower confidence intervals than the PhenoPop method. 
The results are similar for the resistant $GR_{50}$ index.
%Finally, for the resistant $GR_{50}$ value, there is a statistically significant difference in  CI widths among the three different approaches. More precisely, the CI widths of the live cell image method are the tightest among the three, and the CI width of PhenoPop is the widest among the three. 
It is worth mentioning that for at least 28 out of the 30 datasets, the true parameters were located within the confidence intervals for all three methods.

In summary, the end-points and live cell image methods provide a significant improvement in estimator precision over the PhenoPop method for all three parameters, and furthermore, the live cell image method has the best precision out of all three methods.

\subsubsection{Illustrative example with 3 subpopulations}

In this section, we examine a case study involving an artificial tumor with 3 subpopulations.
%the performance of the three methods when inferring parameters for a mixture tumor with 3 subpopulations is investigated.
%Here, the 
The subpopulations are assumed sensitive, moderate, and resistant with respect to the drug, and they are denoted using the subscripts $s,m$, and $r$, respectively. 
Data is generated using a parameter vector $\theta_{BD}(3)$ selected uniformly at random from the ranges in Table \ref{table: modified range}.
Those parameters not listed in Table \ref{table: modified range} are selected as in Table \ref{table: Generating_table_2_pop_illustrative}. 

 \begin{table}[ht]
    \centering
    \begin{tabular}{|c|c|c|c|c|c|c|c|c|}
    \hline
         & $p_{s},p_{m}$&$p_{r}$ &$E_s$& $E_m$ &$E_r$  \\
         \hline
         Range & $[0.167 , 0.333]$ & $1-p_{s} - p_{m}$& $[0.0313 , 0.0625]$  &$[0.25 , 0.375]$&$[1.25,2.5]$\\
         \hline
    \end{tabular}
    \caption{modified range of parameters}
    \label{table: modified range}
\end{table}

\begin{figure}[ht]
    \centering
    \makebox[\textwidth][c]{\includegraphics[scale = 0.2]{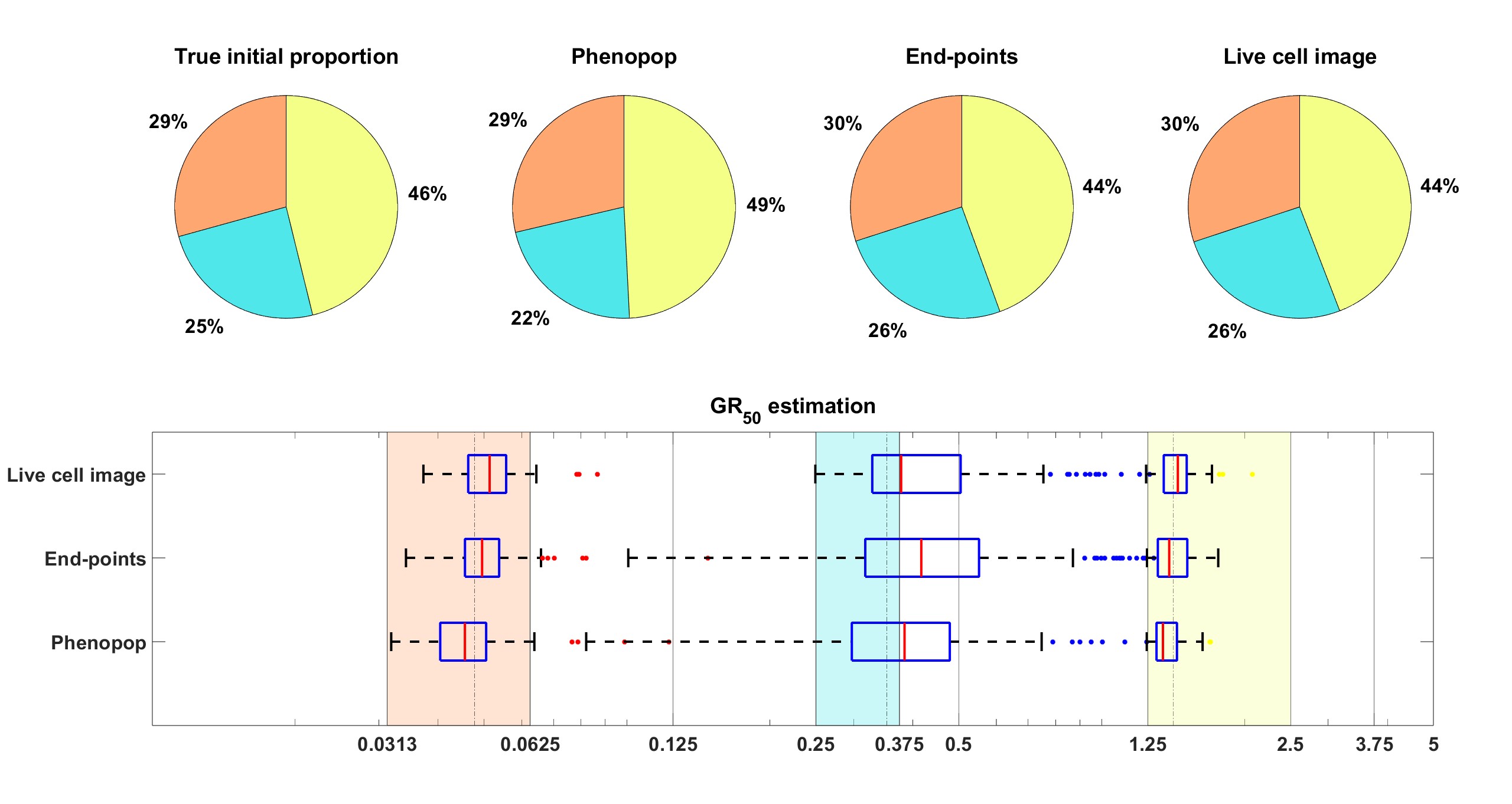}}
    \caption{Estimation of the initial proportion and $GR_{50}$ for 3 subpopulations using the three estimation methods. The parameter vector $\theta_{BD}(3)$ and the observation noise in this example are $p_s = 0.2135,\beta_s = 0.3214,\nu_s = 0.2773,b_s = 0.8782, E_s = 0.0344, m_s = 2.5998, p_m = 0.2718, \beta_m = 0.7334, \nu_m = 0.6776, b_m = 0.8506, E_m = 0.3558, m_m = 4.6055, p_r = 0.5147, \beta_r = 0.0683, \nu_r = 0.0253, b_r = 0.8614, E_r = 1.5764, m_r = 4.4706, c = 9.5209$. The pie chart illustrates the average of all bootstrap estimates for the initial proportion, while the box plot summarizes the distribution of all estimates for the $GR_{50}$'s. The vertical dashed lines in the box plot correspond to the true $GR_{50}$ values employed to generate the data, while the vertical solid lines indicate the concentration levels at which the data were collected. Each color in the plot represents a distinct subpopulation: orange for sensitive, blue for moderate, and yellow for resistant. The shaded areas in the box plot indicate the concentration intervals where the true $GR_{50}$'s are located, and the colored dots mark outliers in the estimation of the $GR_{50}$ for each subpopulation, with red for sensitive, blue for moderate, and yellow for resistant.}
    \label{fig:Illustrative_3}
\end{figure}

Figure \ref{fig:Illustrative_3} shows estimation results for the initial proportions $p_s,p_m,p_r$ and the $GR_{50}$ doses of the three subpopulations.
%In Figure \ref{fig:Illustrative_3}, the \textit{in silico} experiment results of estimating initial proportion and $GR_{50}$ of 3 subpopulations from the mixture data are shown.
Note that the end-points and live cell image methods provide more accurate estimates of the initial proportion for each subpopulation than PhenoPop.
Furthermore, when estimating the $GR_{50}$ for each subpopulation, the inter-quartile range (IQR) of 100 bootstrapped estimates covers the true $GR_{50}$ value for all three methods.
%of all three methods cover the true $GR_{50}$. 
However, the estimation for the $GR_{50}$ of the moderate subpopulation with $E_m = 0.3558$ is less precise than for the other two subpopulations, i.e., the IQR is wider. 
This is likely due to confounding between the moderate subpopulation and the other two subpopulations.

It is worth noting that for the 3 subpopulation example, the number of datapoints is the same as for the 2 subpopulation examples, since only total cell counts are observed at each time point.
%the data set has the same dimension as for the 2 population examples, since we only observe the total number of cells in the population at each time point.
Furthermore, when computing maximum likelihood estimates for 3 subpopulations, we solved each optimization problem the same number of times as for 2 subpopulations.
%and further the same number of optimization runs are used, i.e., the same amount of information and the same effort as the 2 subpopulation experiment. 
%The 3 population model is obviously a more complex model with more parameters, which 
%However, this is of course a more complex model, and this 
%is one possible reason why there is less accuracy and less precision in the estimation.
%the model is of course more complicated when we have more subpopulations. % In fact, we did observe an improvement in precision and accuracy as we increased the number of initial starting points to $300$ for optimization.
Overall, our conclusion is that all three methods can provide reasonable estimates of the true initial proportion and the $GR_{50}$ of each subpopulation for 3 subpopulations. 
However, 
%for 3 subpopulations, 
achieving equivalent levels of accuracy and precision as for 
%the experiments involving 
2 subpopulations may require a greater computational effort or the collection of more data, given that the 3 subpopulation model is more complex and has more parameters.

\subsubsection{Performance in challenging conditions}
\label{sec:Parameter recovering under the extreme case}

In the previous work \cite{kohn2022phenotypic}, three conditions under which the performance of PhenoPop deteriorates were identified:
the case of a large observation noise, a small initial fraction of resistant cells, and similar drug-sensitivity of both subpopulations.
%three challenging scenarios for Phenopop were discovered. 
We now investigate the performance of the end-points and live cell image methods in these conditions and compare to the performance of PhenoPop. \\
%In this section these three settings are investigated: high observation noise, small fraction of resistant cells, and similar sensitivity of both subpopulations.
%Specifically, the performance of PhenoPop, end-points, and live cell image methods are investigated in these three scenarios.\\

\noindent\textbf{Large observation noise:} \\

We first consider the case of large observation noise.
%First consider the scenario where there is a substantial amount of observation noise. 
Note that in the PhenoPop method, the only source of variability in the statistical model is the additive Gaussian noise.
In the end-points and live cell image methods, however, there is an underlying stochastic process governing the population dynamics 
with an added Gaussian noise term.
%in addition to an additive noise term.
%As a result, the end-points and live cell image methods can have difficulty with data with high levels of observation noise.
Thus, whereas PhenoPop deals with high levels of noise by adjusting the variance of the Gaussian term, the two new methods may also try to adjust the subpopulation growth and dose response parameters.
%, as they affect the variance of the model.
This can complicate estimation with the two new methods compared to PhenoPop from data with high levels of noise.
%, whereas PhenoPop can simply adjust the variance level of the additive noise to deal with high levels of observation noise.

\begin{comment}
    Recall that in the data for Figures \ref{fig:Illustrative} - \ref{fig:CI width} $\theta_{BD}(2)$ is selected in the range from Table \ref{table: Generating_table_2_pop_illustrative} which required that $c\in [0,10]$. In addition, the initial cell count is given by $n=1000$. This means the standard deviation is below $1\%$ of the initial cell count. This is in fact lower than the levels used in \cite{kohn2022phenotypic}; this lower level is used because there is already noise from the stochastic population dynamics. 
\end{comment}

\begin{figure}[ht]
    \centering
    \makebox[\textwidth][c]{\includegraphics[scale = 0.2]{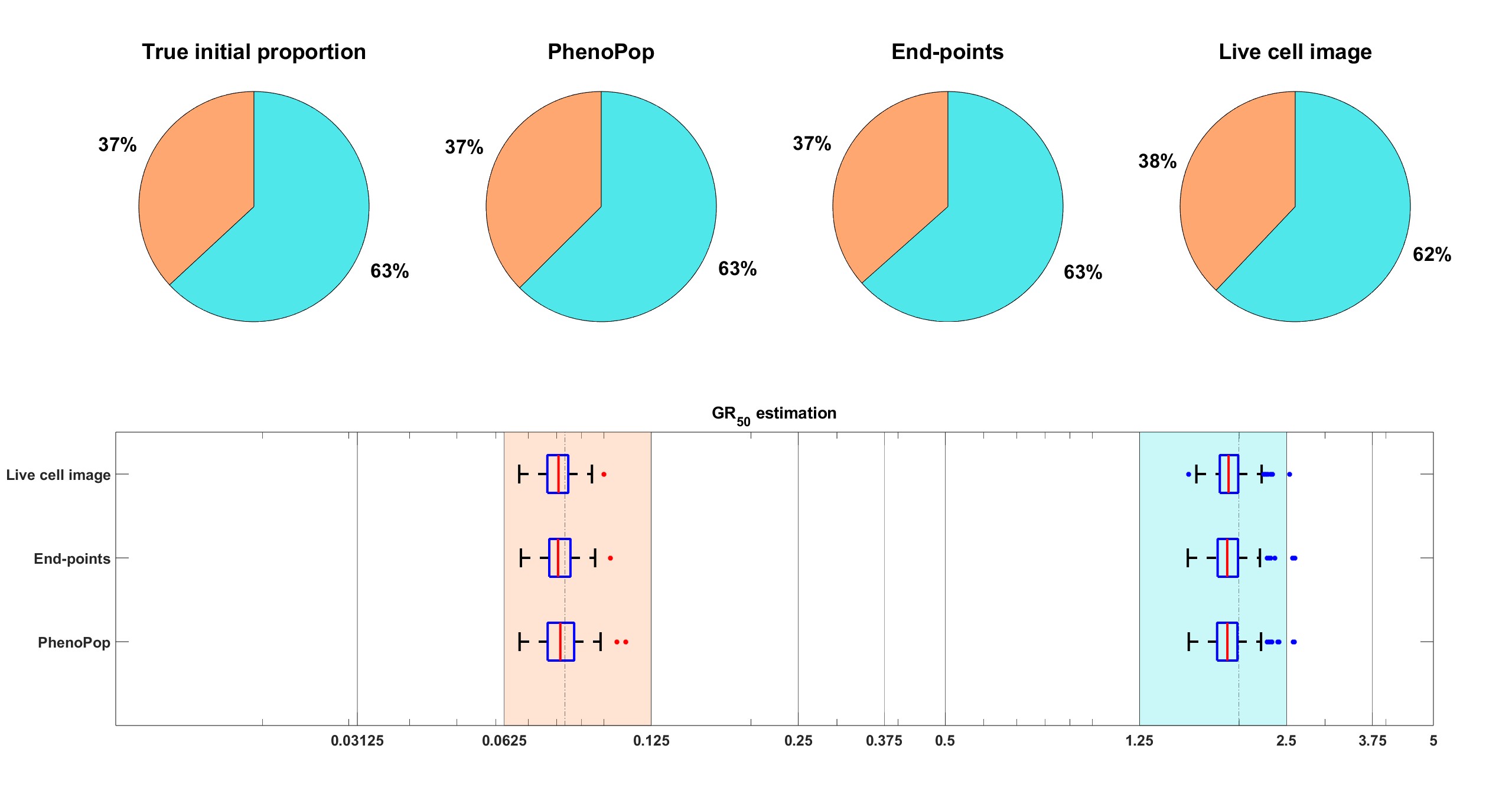}}
    \caption{An illustrative example under the high observation noise scenario, i.e. $c = 500$. The parameter vector $\theta_{BD}(2)$ and the observation noise in this example are $ p_s = 0.3690, \beta_s = 0.4380,\nu_s = 0.3422, b_s = 0.8398, E_s = 0.0813, m_s = 3.9647, p_r = 0.6310, \beta_r = 0.5320, \nu_r = 0.4767, b_r = 0.8674, E_r = 1.9793, m_r = 4.8357, c = 500$. 
    Results are presented as in Figure \ref{fig:Illustrative}. This example demonstrates that all three methods are capable of recovering the initial proportion and $GR_{50}$ even under the high observation noise scenario.}
    \label{fig:ObserationNoise}
\end{figure}

We begin by considering a case study where the noise level is set to $c=500$,
and other parameters are chosen uniformly at random according to Table \ref{table: Generating_table_2_pop_illustrative}.
The results are shown in Figure \ref{fig:ObserationNoise}.
%the results of using the three methods in a higher noise setting are shown. In particular, data is generated with additive noise set at $c=500$, while the rest of the parameters were chosen uniformly at random according to Table \ref{table: Generating_table_2_pop_illustrative}.
%In Figure \ref{fig:ObserationNoise} it can be seen that all 
For each method, the initial proportion $p_s$ is estimated with good accuracy, and the IQR of 100 bootstrap estimates for $GR_{s}$ covers the true value.
%$GR_{s}$ values.
However, compared with the estimation in Figure \ref{fig:Illustrative}, the estimation precision of the end-points method and live cell image method has degraded. In addition,  observe that the IQRs of the three methods have about the same width, which implies the precision advantage observed in Section \ref{sec:Illustrative example} disappears under a very large observation noise. %We did realize that the estimation will get worse as we increase the observation noise, i.e. the estimation for resistant GR-50 in the figure \ref{fig:ObserationNoise} is not as precise as in the figure \ref{fig:Illustrative}.

\begin{figure}[ht]
    \centering
    \includegraphics[width=\linewidth]{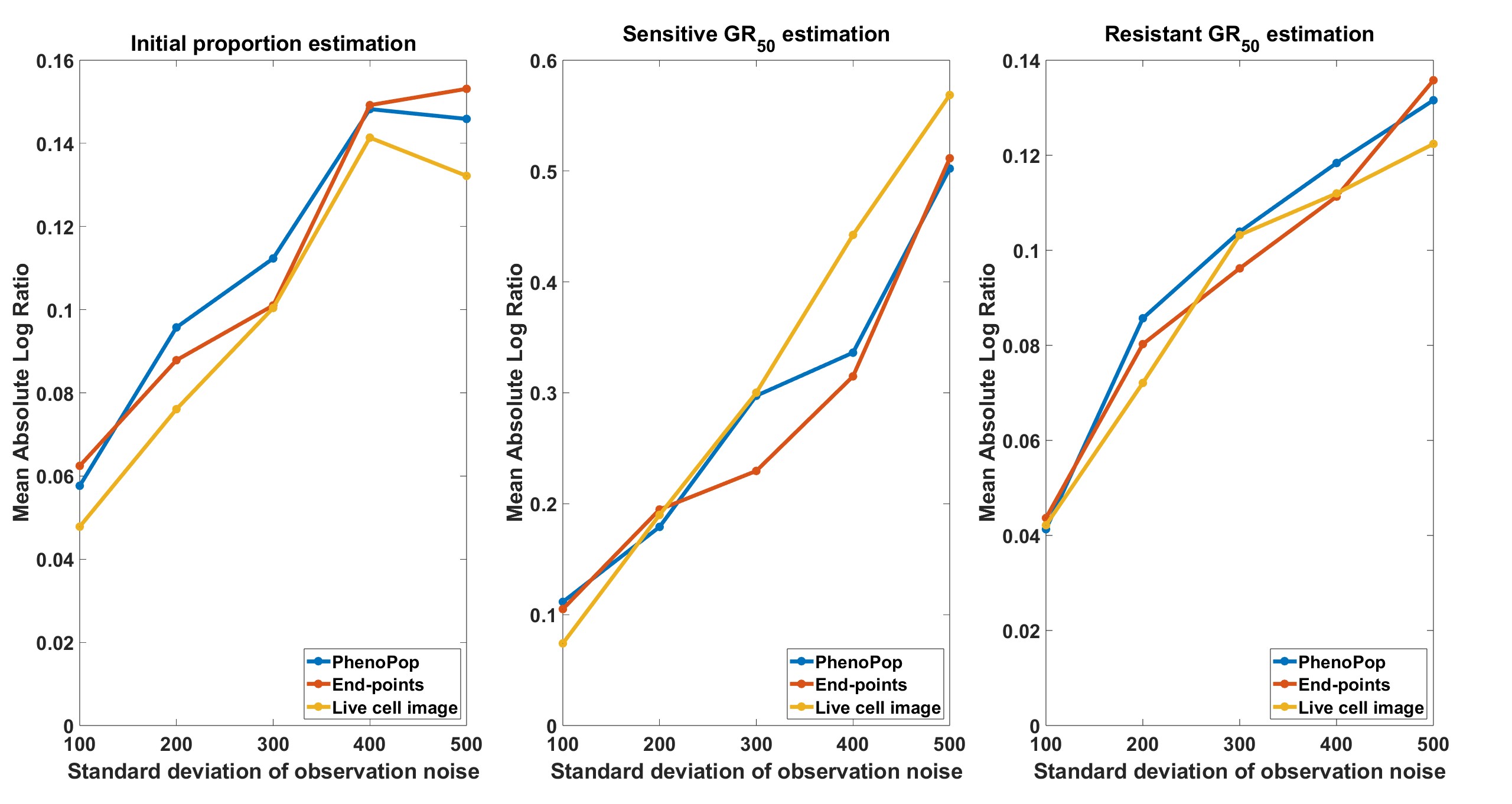}
    \caption{Estimation error of $\{\hat{p}_s,\hat{GR}_s,\hat{GR}_r\}$ with respect to varying standard deviation of observation noise. 
    %estimation accuracy experiment when high observation noise involves.
    The metric of estimation error is the mean absolute log ratio of estimates across 30 simulated datasets, each generated from a distinct parameter set. 
    %The mean absolute log ratio was obtained from 30 independent experiments with 30 randomly generated $\theta_{BD(G)}(2)$ sets. 
    The value of the observation noise parameter, $c$, in these 30 generating parameter sets was assigned to 5 different values in the set $\mathcal{C} = \{100,200,300,400,500\}$ to generate the line plot. Three different line plots correspond to three different methods, as indicated by the figure legends. This figure demonstrates that the estimations of the three methods deteriorate as the level of observation noise increases.}
    \label{fig:sup_noise_quant}
\end{figure}

We next evaluate estimation performance across 30 simulated datasets for each noise value $c\in \mathcal{C} =  \{100,200,300,400,500\}$. 
%It is important to also study the effects of observation noise on estimator accuracy in a quantitative fashion. The results are shown in Figure \ref{fig:sup_noise_quant}. Due to one more dimension, i.e., observation noise, we only plot the average of the absolute log ratio in Figure \ref{fig:sup_noise_quant} rather than the box plots of all results in Figure \ref{fig:Log_Ratio_Accuracy}. 
%Specifically, we again generate 30 random samples of $\theta_{BD(G)}(2)$ uniformly from the range in Table \ref{table: Generating_table_2_pop_illustrative} with different observation noise parameters $c\in \mathcal{C} =  \{100,200,300,400,500\}$. 
Figure \ref{fig:sup_noise_quant} shows the mean absolute log ratio across the 30 datasets for each parameter $\{p_s,GR_{s},GR_{r}\}$, each noise level and each estimation method.
%under the three methods and five noise levels.
As expected, the estimation error 
%In Figure \ref{fig:sup_noise_quant} we can see that the estimation error
increases for all three methods as a function of the observation noise.
In fact, all three methods show
%and that the three methods seem to have 
a similar response to increasing levels of noise.

%Next the behavior of estimator precision, i.e., confidence interval width, is investigated under increasing levels of observation noise. In particular, 30 normalized CI widths were computed at $c=100$ and $c=500$ for all three methods. 
%These results are presented in Figures \ref{fig:CI width 10 noise} and \ref{fig:CI width 50 noise}.

% We generated 30 random samples of $\theta_{BD(G)}(2)$ according to Table \ref{table: Generating_table_2_pop_illustrative}, we then replaced the observation noise parameter $c$ by 100 and 500, i.e. the observation noise has standard deviation $c$ equal to $10\%$ and $50\%$ of initial cell count respectively. 

\begin{figure}[ht]
\begin{subfigure}{.33\textwidth}
  \centering
  % include first image
  \includegraphics[width=\linewidth]{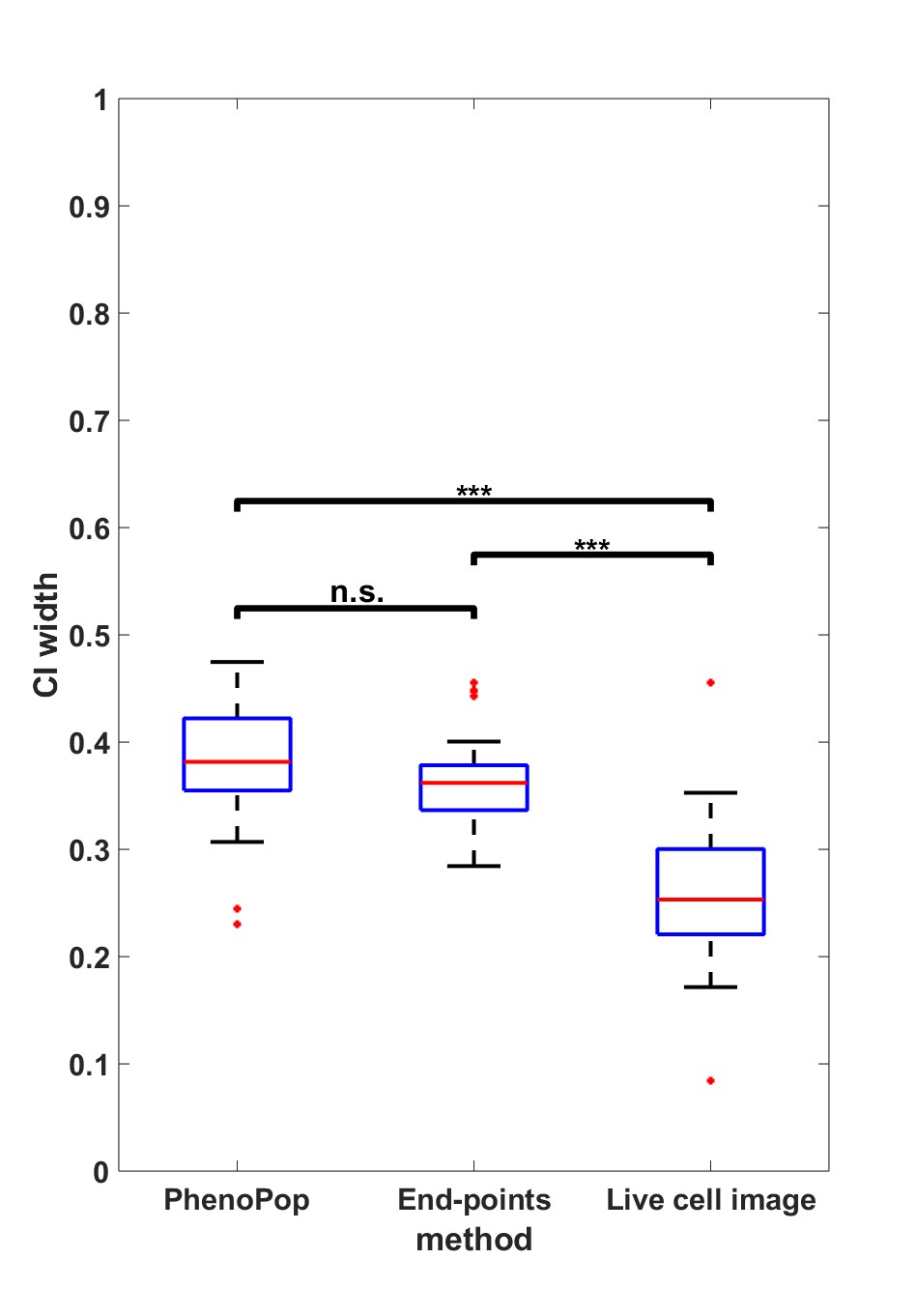}  
  \caption{Initial proportion}
  \label{fig:10_noise_CI_p}
\end{subfigure}
\begin{subfigure}{.33\textwidth}
  \centering
  % include second image
  \includegraphics[width=\linewidth]{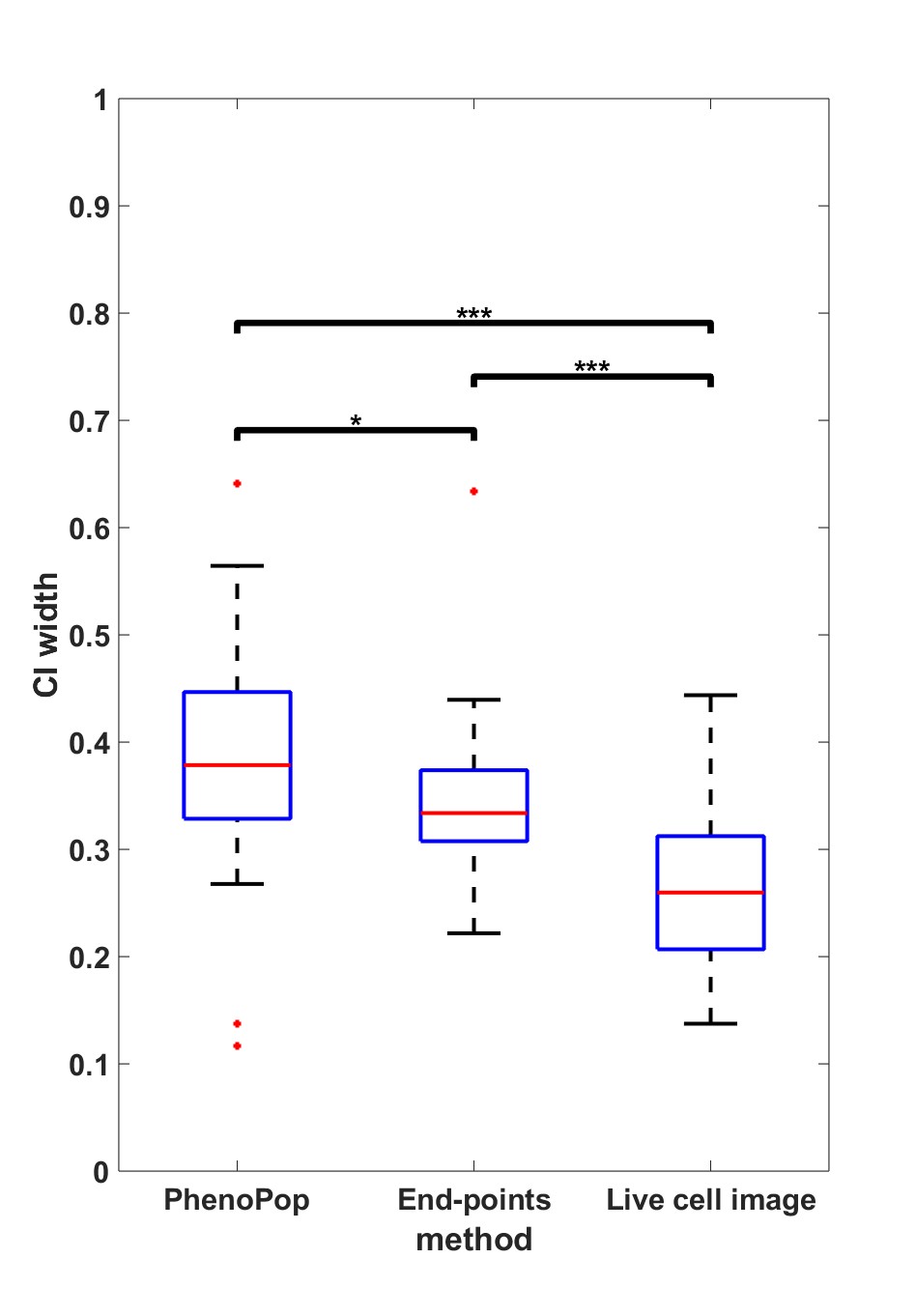}  
  \caption{Sensitive $GR_{50}$}
  \label{fig:10_noise_CI_GR1}
\end{subfigure}
\begin{subfigure}{.33\textwidth}
  \centering
  \includegraphics[width=\linewidth]{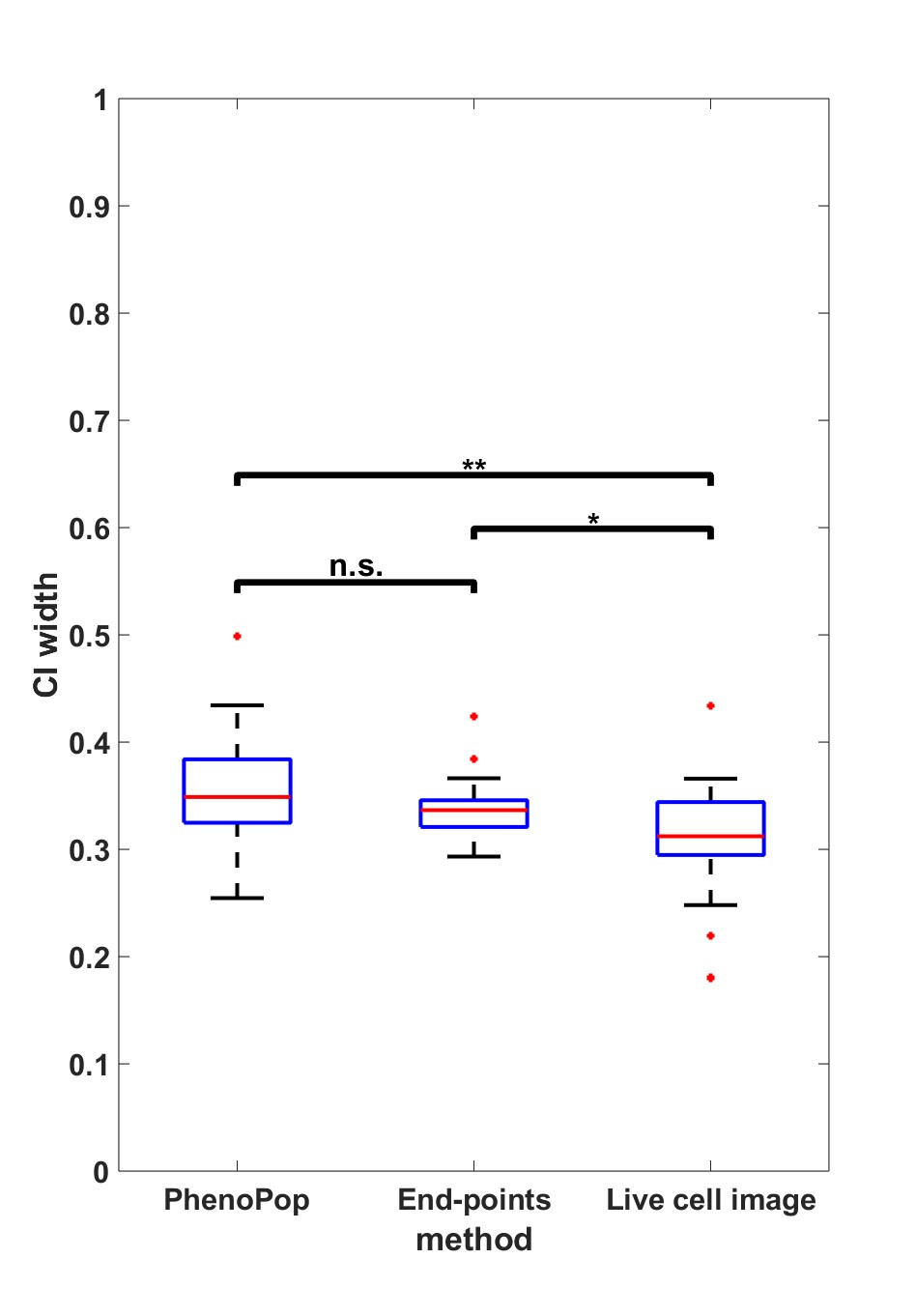}
  \caption{Resistant $GR_{50}$}
  \label{fig:10_noise_CI_GR2}
\end{subfigure}
\caption{Comparison of the normalized CI widths of the three estimators $\{\hat{p}_s, \hat{GR}_s, \hat{GR}_r\}$ using the three different estimation methods, when the observation noise parameter is set to $c=100$. The $y$-axis represents the normalized CI width. The box plot summarizes the results across 30 different datasets. The significance bar indicates the p-values derived from the Wilcoxon rank-sum test, with significance levels denoted as $*** \leq  0.001 \leq  ** \leq   0.01 \leq * \leq 0.05$. This figure demonstrates the advantages of the live cell image method in estimation precision are preserved even when the standard deviation of observation noise is $10\%$ of the initial cell count.}
\label{fig:CI width 10 noise}
\end{figure}

\begin{figure}[ht]
\begin{subfigure}{.33\textwidth}
  \centering
  % include first image
  \includegraphics[width=\linewidth]{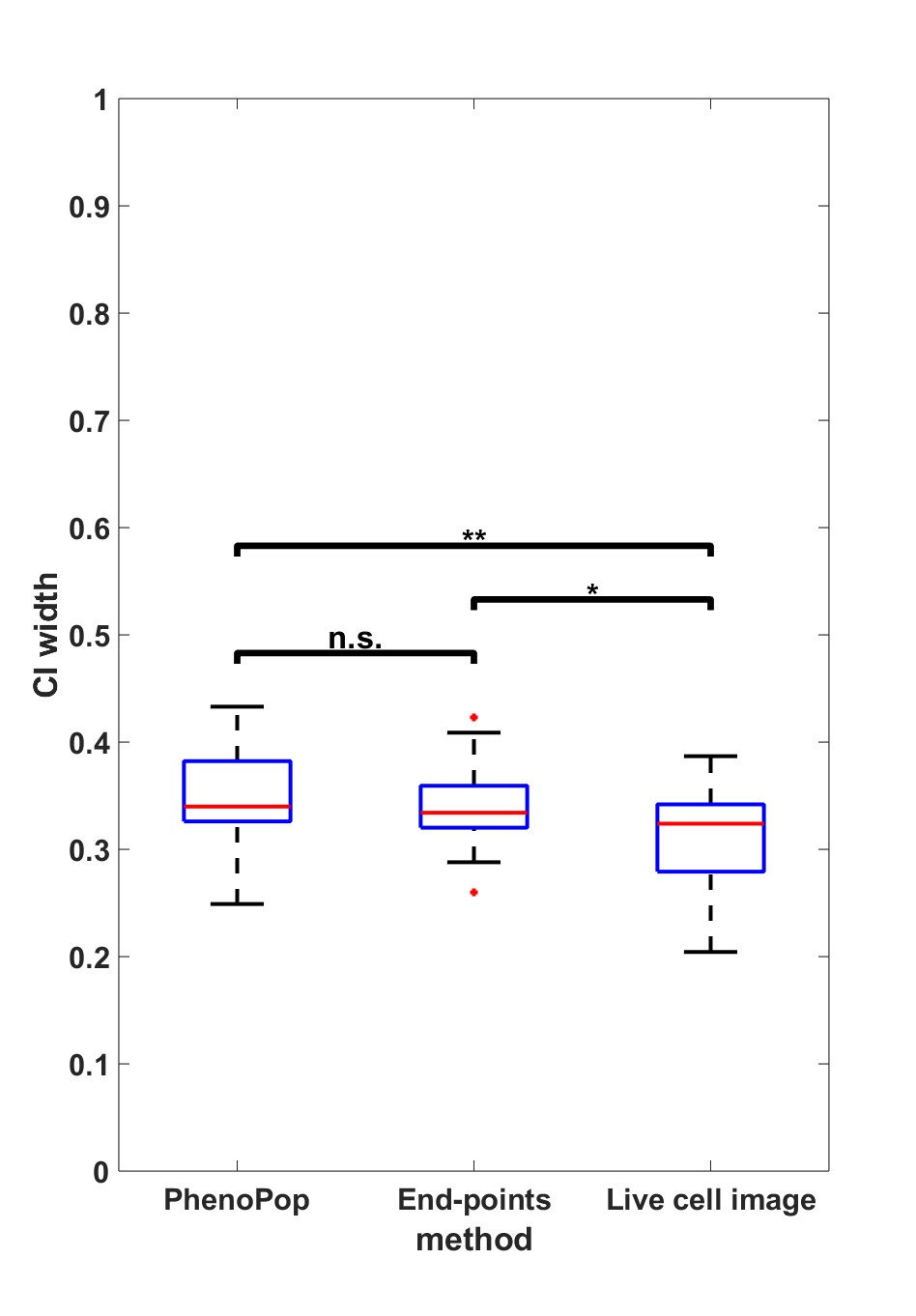}  
  \caption{Initial proportion}
  \label{fig:50_noise_CI_p}
\end{subfigure}
\begin{subfigure}{.33\textwidth}
  \centering
  % include second image
  \includegraphics[width=\linewidth]{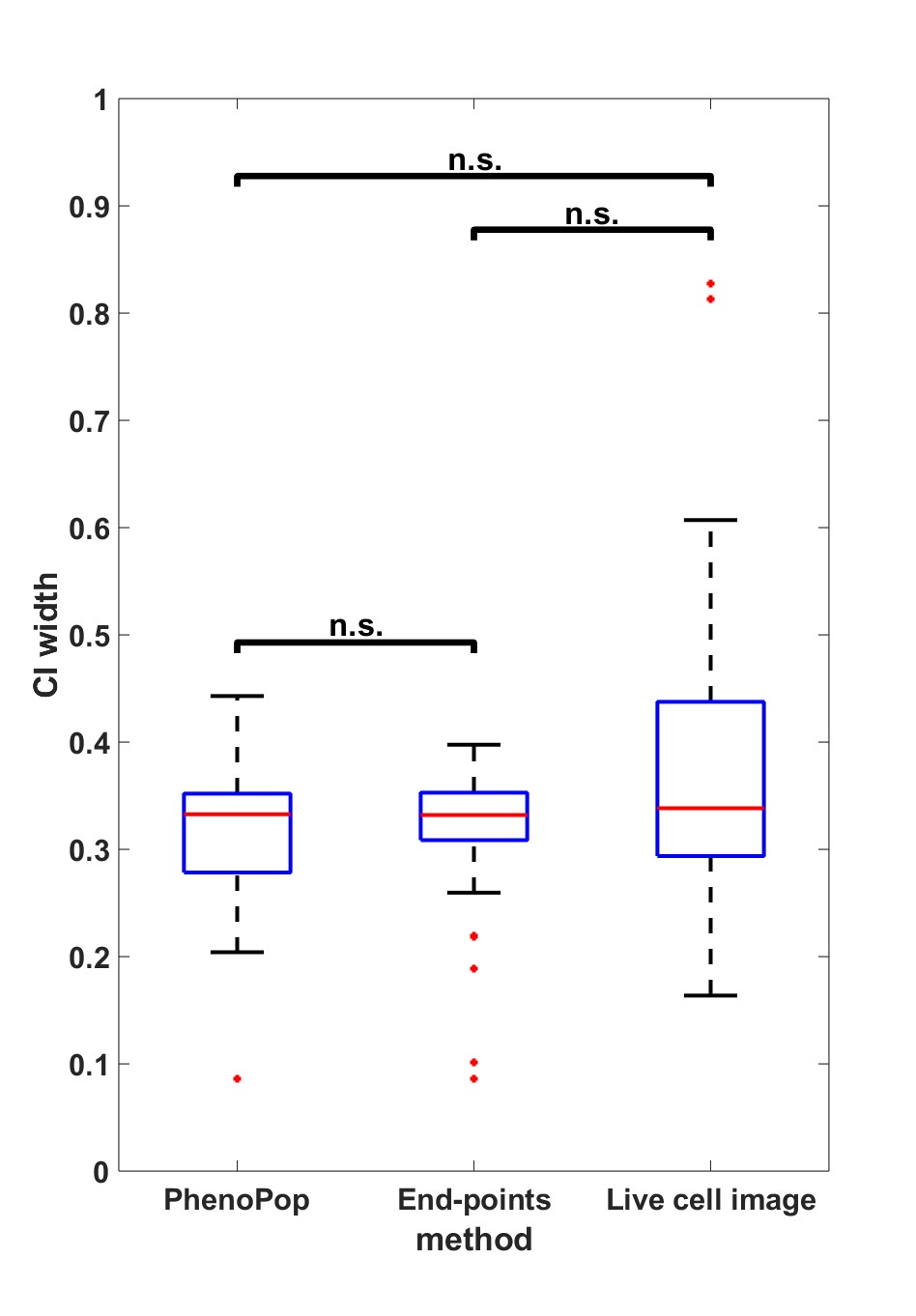}  
  \caption{Sensitive $GR_{50}$}
  \label{fig:50_noise_CI_GR1}
\end{subfigure}
\begin{subfigure}{.33\textwidth}
  \centering
  \includegraphics[width=\linewidth]{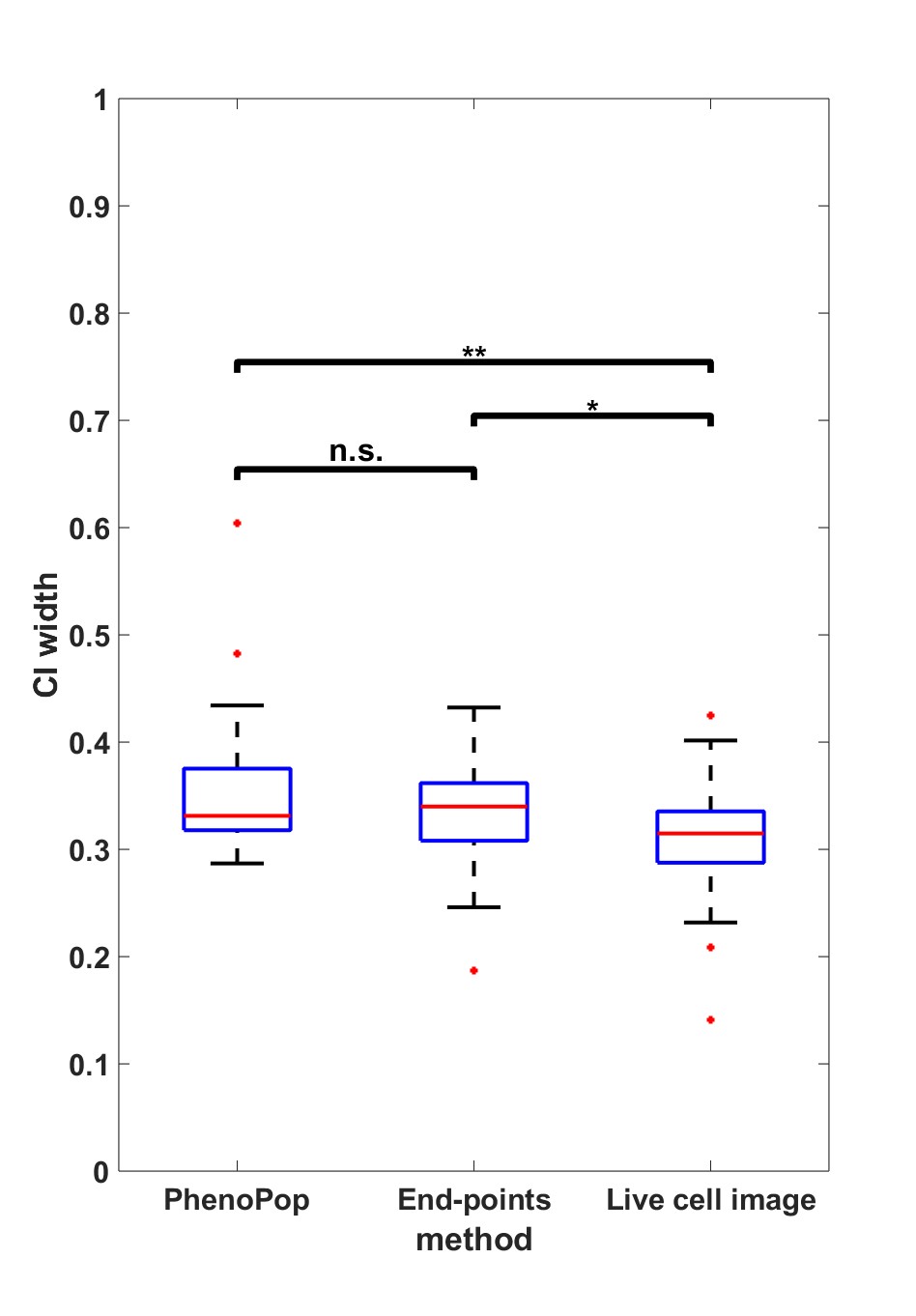}
  \caption{Resistant $GR_{50}$}
  \label{fig:50_noise_CI_GR2}
\end{subfigure}
\caption{Comparison of the normalized CI widths of three estimators $\{\hat{p}_s, \hat{GR}_s, \hat{GR}_r\}$ using the three different estimation methods, when the observation noise parameter is set to $c=500$. Results are presented as in Figure \ref{fig:CI width 10 noise}. This figure demonstrates that the advantages of the live cell image method in estimation precision become less significant as the standard deviation of observation noise increases to $50\%$ of the initial cell count.}
\label{fig:CI width 50 noise}
\end{figure}

We next compare the widths of 95\% confidence intervals for the three parameters under noise levels $c=100$ and $c=500$, using 30 datasets for each noise level.
The results are shown in Figures \ref{fig:CI width 10 noise} and \ref{fig:CI width 50 noise}.
For $c=100$ (Figure \ref{fig:CI width 10 noise}), the precision advantage of the live cell image method over the other two methods is less pronounced than in Figure \ref{fig:CI width}, where $c \in [0,10]$, especially for the resistant $GR_{50}$.
For $c=500$ (Figure \ref{fig:CI width 50 noise}), the advantage disappears for the sensitive $GR_{50}$.
Importantly, however, Figure \ref{fig:CI width 10 noise} shows that the precision advantage of the live cell method is statistically significant for all three parameters $\{p_s,GR_s,GR_r\}$ for an observation noise as large as $10\%$ of the initial cell count. It should be noted that the standard deviation of observation noise reported from common automated and semi-automated cell counting techniques ranges from $1-15\%$ \cite{CADENAHERRERA20159,mumenthaler2011evolutionary}. \\

\noindent\textbf{Small resistant subpopulation:} \\

\begin{figure}[ht]
    \centering
    \makebox[\textwidth][c]{\includegraphics[scale = 0.2]{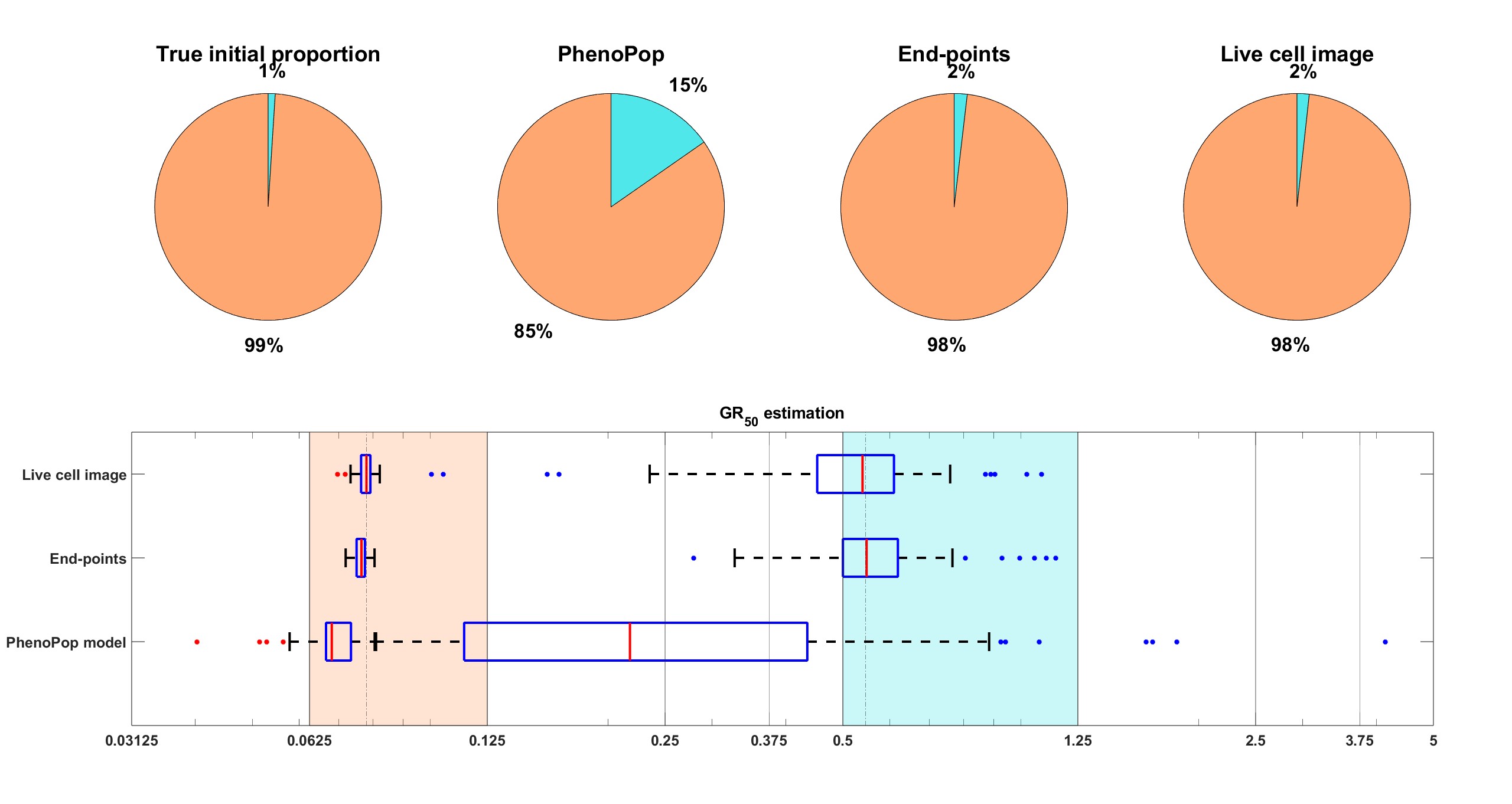}}
    \caption{An illustrative example under the unbalanced initial proportion scenario, i.e. $p_s = 0.99$. The parameter vector $\theta_{BD}(2)$ and the observation noise in this example are $ p_s = 0.9900, \beta_s = 0.4301,\nu_s = 0.4199, b_s = 0.8644, E_s = 0.0768, m_s = 4.3186, p_r = 0.0100, \beta_r = 0.1458, \nu_r = 0.1258, b_r = 0.8565, E_r = 0.5348, m_r = 3.7518, c = 4.8400$. Results are presented as in Figure \ref{fig:Illustrative}. This example demonstrates that our newly proposed model can accurately estimate parameters even when the initial proportion of the resistant subpopulation is negligible, while the PhenoPop method fails to estimate the parameters accurately.}
    \label{fig:High proportion}
\end{figure}

For the datasets investigated in Section \ref{sec:2subpexp}, 
%In the previous examples, 
the initial proportion $p_s$ of sensitive cells was constrained to be in $[0.3,0.5]$. 
%This initial condition ensures generating data with enough information about each subpopulation. 
We now 
%investigate the performance of the estimation methods 
%This constraint is now modified and the performance of the methods is investigated 
%in 
consider
the setting of a small resistant subpopulation.
We begin with a case study in Figure \ref{fig:High proportion}, where $p_s$ is assigned to 0.99, and other parameters are sampled according to Table \ref{table: Generating_table_2_pop_illustrative}.
For both the sensitive and resistant subpopulations, the IQR for the $GR_{50}$ dose under PhenoPop does not cover the true value, whereas the IQR for the live cell image method does. The IQR for the end-points method covers the true resistant $GR_{50}$, but only barely covers the true sensitive $GR_{50}$. In addition, the end-points and live cell image methods have significantly narrower IQRs than PhenoPop.
%For all methods, the sensitive $GR_{50}$ is estimated significantly more accurately than the resistant $GR_{50}$, which reflects the fact that the sensitive population is the dominant subpopulation.
Finally, note that the estimate of the initial proportion of resistant cells is much more accurate for the end-points and live cell image methods.
%than for PhenoPop.
Thus, while PhenoPop provides a reasonable estimate of the sensitive $GR_{50}$, 
which is the dominant subpopulation in this scenario,
inferring the population composition and the $GR_{50}$ for the minority resistant subpopulation requires the use of the more powerful end-points and live cell image methods.

%In particular, now assume initial proportion of the sensitive subpopulation lies in the interval $[0.85, 0.99]$; an example is shown in Figure \ref{fig:High proportion}.

\begin{comment}
    The unbalanced population composition provides a significant amount of information about the dominant subpopulation. As a result, PhenoPop is still able to report information about the dominant subpopulation. However, for the minority subpopulation, there is really limited information available from the data, and in order to infer anything about the minority subpopulation it is necessary to use the more powerful end-points or live cell image methods.
\end{comment}

%In Figure 

%Next a quantitative experiment is performed to validate the observation that the end-points method and the live cell image method outperform PhenoPop in the presence of small resistant subpopulations. The results are shown in Figure \ref{fig:high_prop_Log_Ratio}.

%Specifically, we generated 100 random samples of $\theta_{BD(G)}(2)$ according to Table \ref{table: Generating_table_2_pop_illustrative}, we then discard the component corresponding to the initial proportion of sensitive cells. For each of the randomly sampled $\theta_{BD(G)}(2)$ we consider four possible values for the initial proportion of sensitive cells $\mathcal{P}=\{0.85,0.9, 0.95, 0.99\}$. 
%For each $p\in \mathcal{P}$ and $i\in \{1,\ldots, 100\}$ we compute the absolute log ratio $Er_i^j(p), Er_i^j(GR_s),Er_i^j(GR_r)$ defined in equation \eqref{eq:Log_Ratio}, and then we plot the mean absolute log ratio with respect to the $p\in \mathcal{P}$ in the Figure \ref{fig:high_prop_Log_Ratio}.

\begin{figure}[ht]
    \centering
    \includegraphics[width=\linewidth]{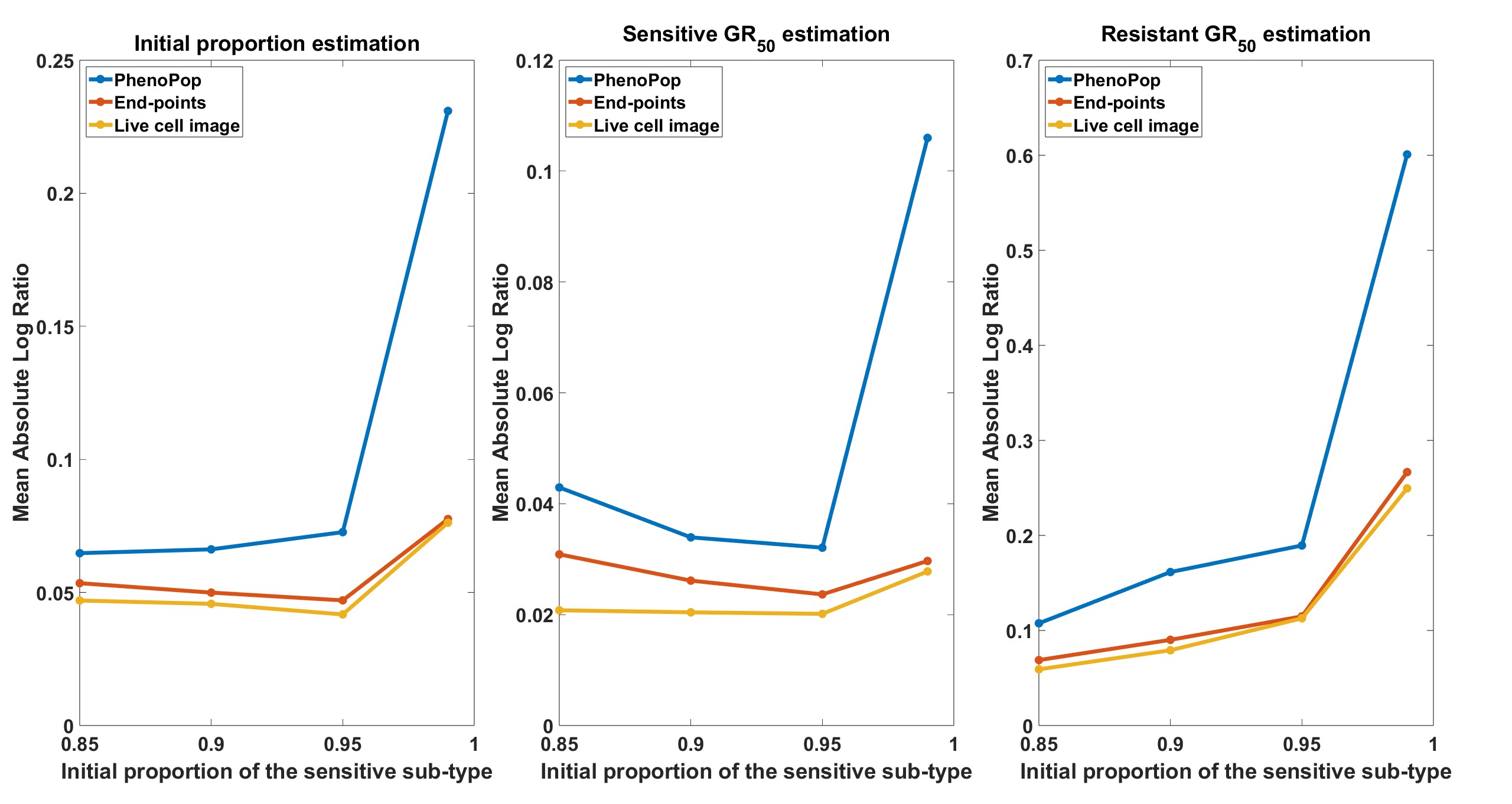}
    \caption{Estimation error of $\{\hat{p}_s,\hat{GR}_s,\hat{GR}_r\}$ with respect  to varying resistant initial proportions.
    %estimation accuracy experiment when limited resistant subpopulation information provided. 
    The metric of estimation error is the mean absolute log ratio across 100 simulated datasets, each generated from a distinct parameter set.  
    %The mean absolute log ratio was obtained from 100 independent experiments with 100 randomly generated $\theta_{BD(G)}(2)$ sets. 
    The value of $p_s$ in these 100 generating parameter sets was assigned to 4 different values in the set $\mathcal{P} = \{0.85,0.90,0.95,0.99\}$ to generate the line plot. Three different line plots correspond to three different methods, as indicated by the figure legends. This figure demonstrates the advantages of estimation accuracy provided by the newly proposed methods when the initial proportion of the resistant subpopulation decreases toward 0.}
    \label{fig:high_prop_Log_Ratio}
\end{figure}

In Figure \ref{fig:high_prop_Log_Ratio}, we show the mean absolute log ratio for each parameter $\{p_s,GR_s,GR_r\}$ across 100 datasets for each $p_s \in \{0.85,0.9,0.95,0.99\}$.
Note that both the end-points and live cell image methods have significantly smaller errors than PhenoPop, and that the difference becomes more pronounced as $p_s$ increases.
%Note that both the end-points method and the live cell image method have smaller errors than the PhenoPop method in the small resistant subpopulation scenario. Furthermore, note that as the initial proportion of sensitive cells increases the accuracy benefit of the live-cell and the end-points methods increases.
Also note that the error in estimating the sensitive $GR_{50}$ is smaller than for the resistant $GR_{50}$, opposite to the results of Figure \ref{fig:Log_Ratio_Accuracy}, where $p_s \in [0.3,0.5]$.
%one may notice that the error of sensitive $GR_{50}$ estimation is now smaller than the error of resistant $GR_{50}$ estimation.
This further reinforces the hypothesis stated in Section \ref{sec:Illustrative example} that the initial proportion of a subpopulation impacts the precision of estimating the $GR_{50}$ for that subpopulation. \\

\noindent \textbf{Similar subpopulation sensitivity:} \\

\begin{comment}
    In the previous experiments, the $GR_{50}$ of the two subpopulations were always at least 3 concentration intervals apart, i.e., there are at least 4 concentration levels between the $GR_{s}$ and $GR_{r}$. In this sub-section, the performance of the three methods is investigated when $GR_s$ and $GR_r$ are allowed to be closer to each other.
Specifically, the $GR_{50}$ of the two subpopulations are located less than 2 concentration intervals apart. These results are presented in Figure \ref{fig:Similar_GR50}, which is the case when $GR_{50}$ of two subpopulations are next to each other. The $\theta_{BD}(2)$ of this experiment was generated from the range given in the Table \ref{table: Generating_table_2_pop_illustrative} except $E_s \in [0.05,0.1],E_r \in [0.15,0.5]$.
\end{comment}

\begin{figure}[ht]
    \centering
    \makebox[\textwidth][c]{\includegraphics[scale = 0.2]{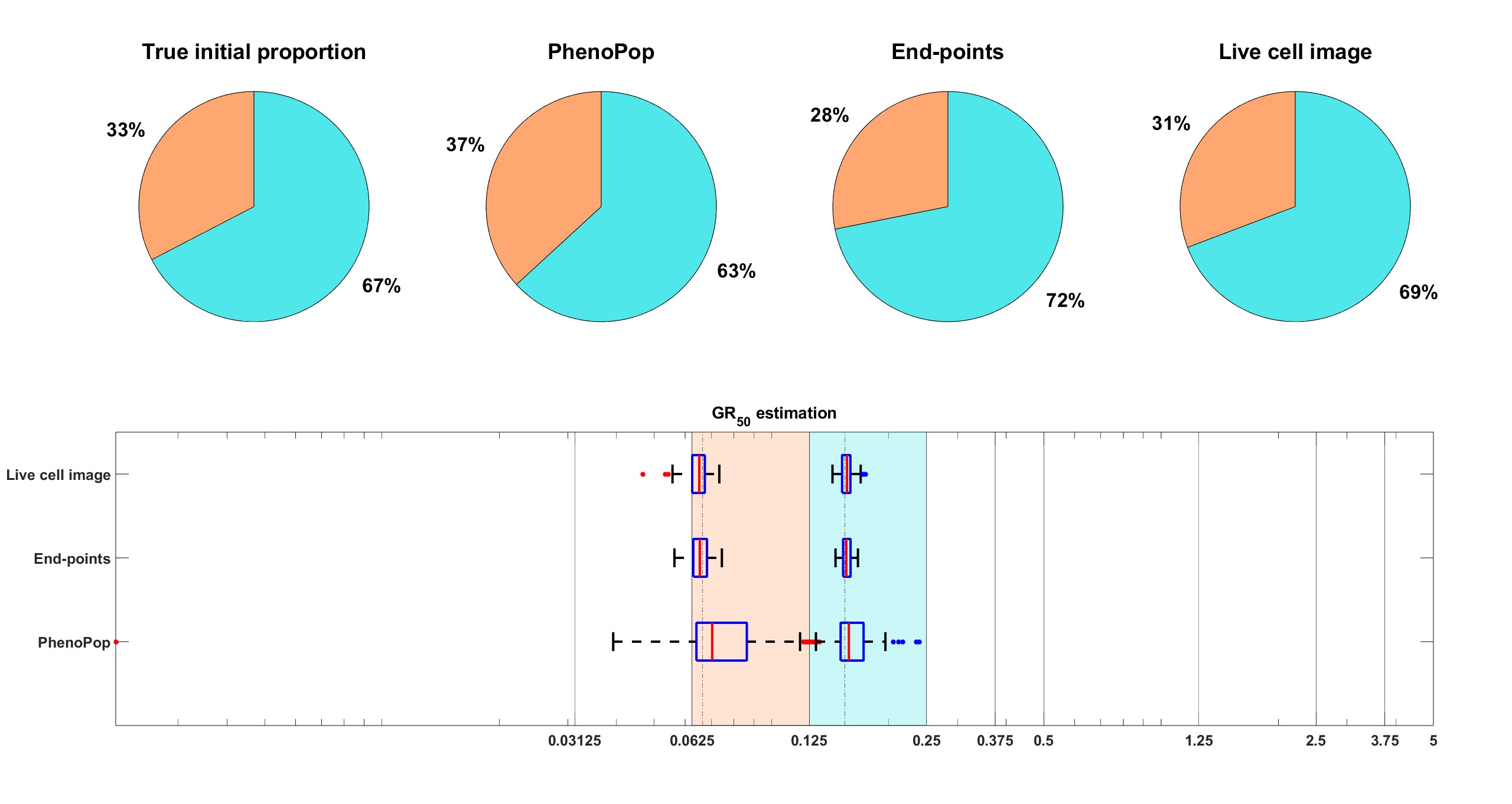}}
    \caption{An illustrative example under the similar subpopulation sensitivity scenario. The parameter vector $\theta_{BD}(2)$ and the observation noise in this example are $ p_s = 0.3263, \beta_s = 0.8896,\nu_s = 0.8215, b_s = 0.8820, E_s = 0.0654, m_s = 3.8539, p_r = 0.6737, \beta_r = 0.0925, \nu_r = 0.0661, b_r = 0.8171, E_r = 0.1500, m_r = 3.6015, c = 7.6660$. Results are presented as in Figure \ref{fig:Illustrative}. This example demonstrates that all three methods are capable of recovering the initial proportion and $GR_{50}$ even when two subpopulations have similar drug sensitivity, while the newly proposed methods exhibit superior estimation precision compared to the PhenoPop method.}
    \label{fig:Similar_GR50}
\end{figure}

\begin{figure}[ht]
    \centering
    \includegraphics[width=\textwidth]{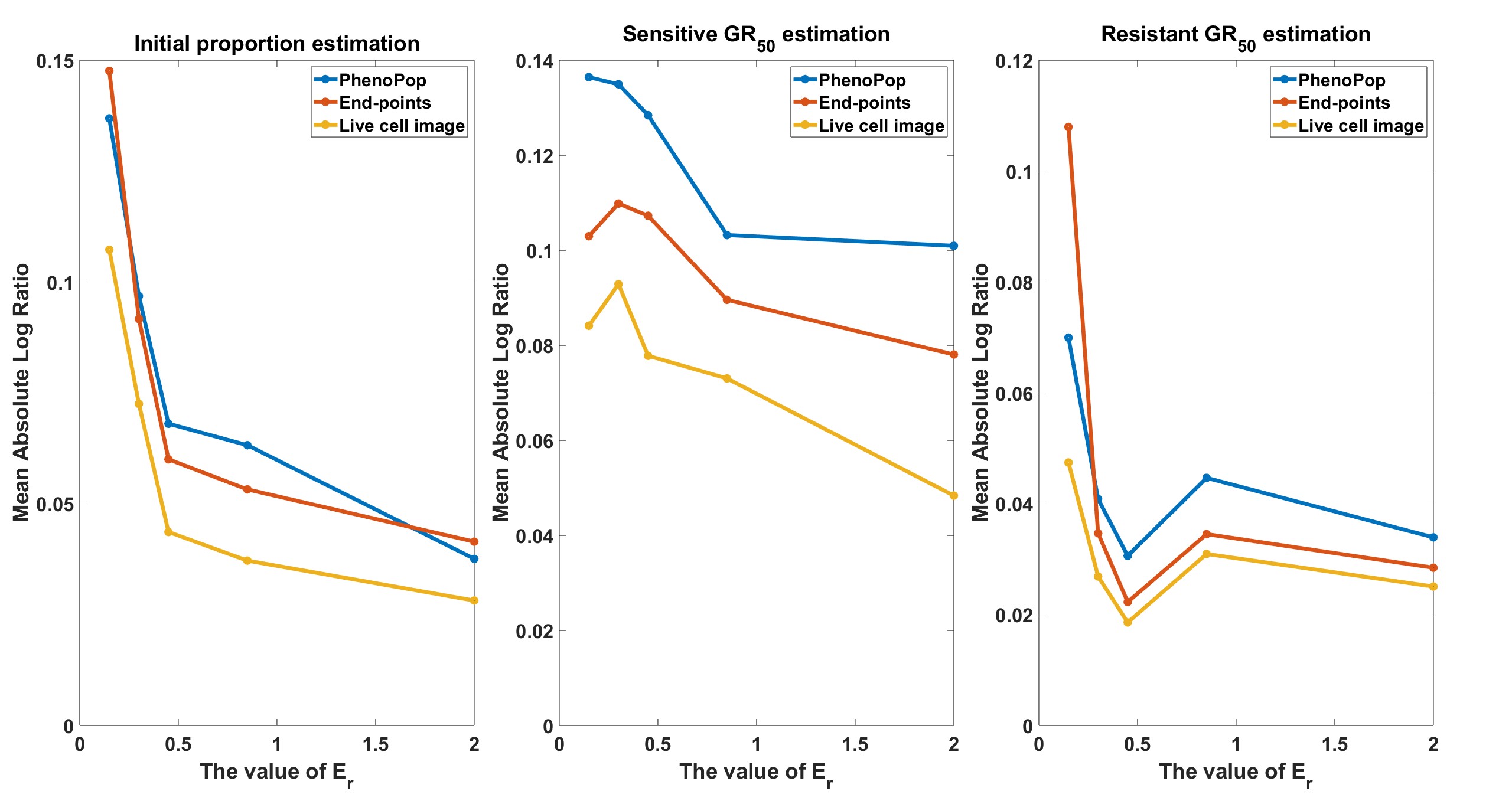}
    \caption{Estimation error of $\{\hat{p}_s,\hat{GR}_s,\hat{GR}_r\}$ with respect to varying similarity between subpopulation drug sensitivities.
    %high observation noise involves.
    The metric of estimation error is the mean absolute log ratio across 80 simulated datasets, each generated from a distinct parameter set. 
    %The mean absolute log ratio was obtained from 80 independent experiments with 80 randomly generated $\theta_{BD(G)}(2)$ sets. 
    The value of $E_r$ in these 80 generating parameter sets was assigned to 5 different values in the set $\mathcal{E} = \{0.15,0.3,0.45,0.85,2\}$ to generate the line plot. Three different line plots correspond to three different methods, as indicated by the figure legends. This figure demonstrates that the estimation accuracy of the three methods improves as the discrepancy of drug sensitivity between the two subpopulations increases, with the live cell image method exhibiting the smallest average error among the three methods.}
    \label{fig:Close_GR_Log_Ratio}
\end{figure}

For the datasets investigated in Section \ref{sec:2subpexp}, 
%In the previous examples, 
the $GR_{50}$'s for the two subpopulations were assumed to be significantly different.
We now consider the case where the two $GR_{50}$'s are similar.
Figure \ref{fig:Similar_GR50} shows the results of a case study where $E_s \in [0.05,0.1]$, $E_r = 0.15$, and other parameters are selected according to Table \ref{table: Generating_table_2_pop_illustrative}.
%In Figure \ref{fig:Similar_GR50}, note 
Note that all three methods successfully recover the parameters $\{p_s,GR_s,GR_r\}$, where the IQRs for the live cell image method are significantly narrower than for PhenoPop. For brevity, we omit the plots that depict the statistical comparison of confidence interval widths. In Figure \ref{fig:Close_GR_Log_Ratio}, we perform estimation across 80 datasets for each $E_r \in \{0.15,0.3,0.45,0.85,2.0\}$, with other parameters sampled from Table \ref{table: Generating_table_2_pop_illustrative}, including $E_s \in [0.05,0.1]$.
As expected, the accuracy in estimating the parameters $\{p_s,GR_s,GR_r\}$ improves as the sensitive $GR_{50}$ and resistant $GR_{50}$ become more different.
%, which is a reasonable observation because two subpopulations are getting more distinguishable. 
We note however that the live cell image method has the lowest mean error when estimating the parameters, with all three methods showing similar degradation as the two subpopulations become more phenotypically similar.

% Besides, we also notice that the Log Ratio of estimating resistant $GR50$ is increasing when $E_2>0.5$. We consider this phenomenon may not be due to the noise of the experiment. One possible reason is that the concentration interval we select may not cover enough information about the resistant $GR50$ as we increase the value of $E_2$. Nevertheless, we can observe that the Log Ratio of resistant $GR50$ estimation is around $0.03$ when $E_2 = 0.45$, which is the lowest among all three parameters estimation. Thus, we suppose the estimation of resistant $GR50$ may reach its optimal precision when $E_2 = 0.45$ under the current concentration design. In general, we consider the error in this scenario is acceptable, i.e. the largest mean error is around $0.2$. This result means in most of the time, three models can successfully identify the correct concentration interval where the true $GR50$ locate from synthetic data as we observed in Figure \ref{fig:Similar_GR50}. 

\subsection{Application to \textit{in vitro} data} \label{sec:invitro}

We conclude by evaluating the performance of our two new methods on 
{\em in vitro} experimental data.
The data consists of different mixtures of imatinib sensitive and resistant \textit{Ba/F3} cells.
In the experiments, cells were exposed to 11 different concentrations of imatinib and they were observed at 14 different time points.
For each drug concentration, 14 independent replicates were performed starting with roughly 1000 cells.
Cell counts were obtained using 
a live-cell imaging technique.
Four  datasets were produced with different starting ratios between sensitive and resistant cells: $1:1$, $1:2$, $2:1$ and $4:1$. 
These datasets are denoted BF11, BF12, BF21 and BF41, respectively.
See \cite{kohn2022phenotypic} for further details on the experimental methods for generating the data.

In \cite{kohn2022phenotypic}, we showed that the PhenoPop method can accurately identify the initial proportion of sensitive cells and both subpopulations' $GR_{50}$ indices from the datasets. 
%Concentration intervals here refer to a discretization of the possible range for the $GR_{50}$s, as is further explained in \cite{kohn2022phenotypic}.
Here, we apply the two new estimation methods to the datasets and compare how well the models fit the data.
%Therefore, in this paper, the focus is on comparing all three methods. 
%The performance of algorithms 
Model fits are assessed using the Akaike Information Criterion (AIC), which for a statistical model with parameters $\theta$ and likelihood function $\mathcal{L}(\theta|\mathbf{x})$ is given by
\begin{align*}
    AIC &= -2 \log(\mathcal{L}(\theta^*|\mathbf{x})) + 2|\theta^*|.
\end{align*}
Here, $\theta^*$ is the maximum likelihood esitmate and $|v|$ is the cardinality of the vector $v$. When comparing the three methods, the one with the lowest AIC is preferred. 

%\textbf{\textit{Ba/F3} data feasible region for optimization}

\begin{table}[ht]
    \centering
    \begin{tabular}{||c|c|c|c||}
         DATA & PP(AIC) & EP(AIC) & LC(AIC)   \\
         BF11& 28502 & 26300 & \textbf{25294} \\
         BF12& 30485 & \textbf{26816} & 27311 \\
         BF21& 27928 & \textbf{24064} & 24182 \\
         BF41& 28912 & \textbf{24066} & 24574 \\
    \end{tabular}
    \caption{AIC scores of three methods: PhenoPop method(PP), end-points method(EP), and live cell image method(LC) for the four experimental datasets BF11, BF12, BF21 and BF41. 
    %These scores are obtained from optimization with 500 randomly chosen feasible initial points. The feasible region is described in Table \ref{table:OptimizationConstriantsInVitro}.
    }
    \label{table:AICofBa/F3}
\end{table}

Results are shown in Table \ref{table:AICofBa/F3}.
The AIC values of the end-points method (EP) and live cell image method (LC) 
%in Table \ref{table:AICofBa/F3} 
are clearly lower than for the PhenoPop method (PP), indicating that the two new methods are superior for fitting the experimental datasets. As discussed in Section \ref{sec:Model}, the newly proposed methods have more sophisticated variance structures, which is likely the reason why they are able to provide a better fit to the datasets. Finally, we note that the end-points method has superior AIC scores to the live cell image method for three out of four of the experimental conditions. 
Therefore, it is not clear which of these two methods is more appropriate for fitting these datasets. 
%{\color{blue} This is a little bit of a weak conclusion.}

\section{Discussion}
\label{sec:Discussion}

In this work, we have proposed two methods for analyzing data from heterogeneous cell mixtures. In particular, we are interested in the setting where a mixture of at least two distinct cell subpopulations is exposed to a given drug at various concentrations. We then use the dose response curve of the composite population to learn about the two subpopulations. In particular, we are interested in estimates of the different subpopulations' initial prevalence and also their distinct dose response curves. The challenge of this problem is that we do not observe direct information about the subpopulations, but instead only information about the dose response of the composite population. 

This work is an extension of our prior work in \cite{kohn2022phenotypic}. The novelty of the current work is that we introduce a more realistic variance structure to our statistical model. We create a new variance structure by building our model using linear birth-death processes. In particular, we model each subpopulation as a linear birth-death process with a unique birth rate and a unique dose-dependent death rate. The dose dependence of the death rate is captured using a 3-parameter Hill function. Our observed process is then a sum of independent birth-death processes. Our goal is then to estimate the initial proportion of the subpopulations, as well as their birth rates and the parameters governing the dose response in their death rates.

Counting cells in \textit{in vitro} experiments can generally be conducted in one of two fashions. In the first approach, cell numbers can only be estimated at the end of the experiment because the mechanism for estimating cell numbers requires killing the cells. In the second approach, cells are counted via live imaging techniques and the cells can be counted at multiple time points. When dealing with multiple time point data from cells collected via the first approach we can assume that observations at different time points are independent because they are the result of different experiments. However, when dealing with data from the second approach we can no longer make that assumption because the cell counts at different time points are from the same population and there is a positive correlation between those measurements. As a result of this differing structure we develop two methods, one that assumes independent observations at each time point, and one that assumes all the time points for a given dose are correlated. Evaluating the likelihood function under the second approach is not trivial at first glance since it requires evaluating the likelihood function of a sample path of a non-Markovian process (the total cell count). We are able to get around this difficulty by using a central limit theorem argument to approximate the exact likelihood function with a Gaussian likelihood.

In this work we compared three different methods: PhenoPop method from \cite{kohn2022phenotypic}, end-points method (assumes measurements are independent in time), and live cell image method (assumes time correlations). We first performed this comparison using simulated data. We generated our data by simulating linear birth-death processes and then adding independent Gaussian noise terms to the simulations. 
We mainly focused on a mixture of two supbopulations,
and 
%Throughout our tests 
we were interested in estimating three features of the mixed population: initial proportion of sensitive cells, $GR_{50}$ of the sensitive cells, and $GR_{50}$ of the resistant cells.
Our first test for the simulated data was to look at confidence interval widths as a measure of estimator precision. In this study, we found that the live cell image method had significantly narrower confidence intervals than the other methods for estimating all three features. We next investigated the performance of our three methods in the setting of small resistant subpopulations, where less than 15\% of initial cells are resistant. We found that in this small resistant fraction setting the live cell image method provides a significant improvement in accuracy over the original PhenoPop method. Furthermore, this improvement increases as the initial fraction of resistant cells goes to zero. We also compared the performance of the methods for simulations with increased levels of additive noise and subpopulations with similar dose response curves. In the scenario of subpopulations with similar dose response curves, we found that the live cell image method has the lowest mean error among the three methods.
For increasing additive noise, all three methods perform similarly in terms of estimation accuracy.
However, the live cell image method maintains its precision advantage over the other two methods for an observation noise of 10\% of the initial cell count, while the advantage disappears for a 50\% noise level.
%In general, we observe that the live cell image method maintains its advantage in estimation precision over the other methods in most of the simulations we conducted.
%However, in the simulation with increasing additive noise, we observed that all three methods perform similarly in terms of estimation accuracy. Nevertheless, we note that the live cell image method maintains its advantage in estimation precision across most of the simulations we conducted.

% In these simulations, we found that the two new methods perform comparably with the original PhenoPop method. 

We finally compared the three methods using \textit{in vitro} data. In particular, we used data from our previous work \cite{kohn2022phenotypic} that considered different seeding mixtures of imatinib sensitive and resistant tumor cells. We then used all three methods to fit this data and used AIC as a model selection tool. We found that live cell image and end-points methods had significantly better scores than PhenoPop for all four initial mixtures studied. Interestingly the end-points method had lower AIC scores for three out of the four mixtures studied even though this data was generated using live-cell imaging techniques. 

In our statistical model, there are several important features of cell biology that we have left out.  For example, one type of cell may transition to another type of cell via a phenotypic switching mechanism (see e.g., \cite{gunnarsson2020understanding,gunnarsson2023statistical}). We believe that our current methods should be able to handle this type of switching with little modification since the underlying stochastic model will be very similar, i.e., a multi-type branching process. Another way the cell types can interact is via competition for scarce resources as the populations approach their carrying capacity. These types of interactions will require new statistical models since the underlying stochastic processes will no longer be linear birth-death processes. Another interesting direction of future work is to quantify the limits of when we can identify distinct subpopulations. For example, if the resistant subpopulation is present at fraction $\epsilon$, what observation set would allow us to identify the presence of this subpopulation? Finally our stochastic model assumes that the time between cell divisions is exponential, but this is of course a great simplifcation. At the cost of a more complex model it would be possible to incorporate states for the different stages of the cell cycle. We leave this open as a question for future investigation.

\section{Appendix}

\subsection{Details of the numerical experiments}
%{Environment setting of the numerical experiment}
%{\color{red} Can this entire section be moved to Appendix?}
\label{appx: Environment setting}
% We first define some of the basic algorithms and formulas used in the following numerical results.
We define some of the basic algorithms and formulas used in the numerical results.

\subsubsection*{Generation of simulated data}

To simulate data, the parameter set $\theta_{BD}(S)$ (as defined in \eqref{eq:BD_param_def}) is selected uniformly at random from a subset of the parameter space given in Table \ref{table: Generating_table_2_pop_illustrative}. Note that one can obtain the generating parameter set $\theta_{PP}(S)$ (defined in \eqref{eq:PP_param_def})from $\theta_{BD}(S)$ directly by setting $\alpha_i=\beta_i-\nu_i$ for each subpopulation. Based on the parameter set $\theta_{BD}(S)$, we simulated data is generated according to the statistical model specified in equation \eqref{eq:stat_model}. Note that data is collected from the simulation continuously during the course of the experiment to replicate the live-cell imaging experiments.

\begin{comment}
    \subsubsection*{Experimental data}

Experimental dataset were obtained from drug-response assays on admixtures of imatinib-sensitive and -resistant Ba/F3 cells.  These assays were performed using the live-cell imaging technique, see the experimental methods section of \cite{kohn2022phenotypic}. 
\end{comment}

\subsubsection*{Maximum likelihood estimation (MLE)}

The maximum likelihood estimation was conducted by minimizing the negative log-likelihood, subject to constraints that were placed on the range of each parameter. The optimization process to find the minimum point was based on the MATLAB Optimization Toolbox \cite{MatlabOTB} function fmincon with sequential quadratic programming (sqp) solver. Due to the non-convexity of the negative log-likelihood function, we performed the optimization starting from 100 uniformly sampled initial points within a feasible region. The feasible region sets limitations on the parameters based on prior knowledge about them. For simulation studies, the feasible region is given by Table \ref{table:OptimizationConstraintsInSilico}, and for the \textit{in vitro} data the feasible region is specified by Table \ref{table:OptimizationConstriantsInVitro}. Among all the resulting local optima, the parameter set with the lowest negative log-likelihood as the estimated result. 

\subsubsection*{Bootstrapping}

In the simulated experiments,  bootstrapping is used to quantify the uncertainty in the MLE estimator. In particular, $20$ independent replicates of data measured at 11 concentration values $\mathcal{D}$ and 13 time points $\mathcal{T}$ are generated from the parameters $\theta_{BD}(S)$ at the beginning of the experiment. Then bootstrapping is employed to randomly re-sample $13$ replicates from those $20$ replicates with replacement $100$ times. With 13 randomly sampled replicates it is possible to create an MLE  for the parameter set $\theta_{BD}$. Since there are now 100 MLE's for $\theta_{BD}$ it is possible to construct confidence intervals as well by using the empirical quantiles of the estimators.

\subsubsection*{$GR_{50}$}

Our goal is to estimate the number of subpopulations, initial mixture proportion $p_i$, and the drug sensitivity of each cellular subpopulation. The $GR_{50}$, introduced in \cite{Hafner2016}, is a summary metric of drug-sensitivity. It is defined as the concentration at which a drug's effect on cell growth is half the observed effect. Note that at the maximum concentration level, the drug may not reach its theoretical maximum effect. 

In the context of the model, the $GR_{50}$ can be defined as below.
Denote the maximum dosage applied as $d_m$, and define the half-maximum effect for subpopulation $i$ as $r_i = (\nu_i(d_m) + \nu_i(0))/2$. The explicit formula for the $GR_{50}$ is then for subpopulation $i$:
\[GR_{50} = E_i\left(\frac{1 - e^{\nu_i - r_i}}{e^{\nu_i - r_i} - b_i} \right)^{1/m_i}\]

When $S=2$, we will denote the higher $GR_{50}$ as either the resistant $GR_{50}$ or $GR_{r}$, and the lower $GR_{50}$ as either the sensitive $GR_{50}$ or $GR_{s}$. In this setting, parameters for the sensitive subpopulation and the resistant subpopulation, respectively, are denoted by subscripts $s$ and $r$, e.g. $E_s$ and $E_r$.

\subsubsection*{Initial conditions.}
The initial number of cells is  set as $n = 1000$, and  the initial size of each subpopulation is set by rounding $n p_i$ to the nearest integer for subpopulation $i$. The following drug concentration levels are used
$$
\mathcal{D} = [0, 0.0313, 0.0625, 0.1250, 0.2500, 0.3750, 0.5, 1.25, 2.5, 3.75, 5]
$$
and we collect the cell count data at the time points: 
$$
\mathcal{T}= [0, 3, 6, 9, 12, 15, 18, 21, 24, 27, 30, 33, 36].
$$
For these specific concentration levels and time points, we have chosen the threshold values of $T_L = 21$ and $D_L = 1$ in the PhenoPop model.

\subsubsection*{Optimization feasible region}
When performing numerical optimization the parameters are restricted to a physically realistic region.
Unless otherwise noted, the optimization  was performed using 100 uniformly sampled initial points from Table \ref{table:OptimizationConstraintsInSilico}.

\begin{table}[ht]
    \centering
    \begin{tabular}{|c|c|c|c|c|c|c|c|c|c|}
    \hline
         & $p_s$ &$p_r$& $\beta_{s,r}$ & $\nu_{s,r}$ & $b_{s,r}$ & $E_{s,r}$ & $m_{s,r}$ & $\sigma_L,\sigma_H$ & $c$  \\
         \hline
         Range & $[0,0.5]$ &$1-p_s$& $[0,1]$ & $[\beta-0.1,\beta]$ & $[0.27,1]$ & $[0,10]$& $[0,10]$ & $[0,2500]$ & $[0,10]$ \\
         \hline
    \end{tabular}
    \caption{Feasible interval for each parameter.}
    \label{table:OptimizationConstraintsInSilico}
\end{table}

\subsection{Estimation on {\em in vitro} experimental data}

When solving the maximum likelihood optimization problems for the {\em in vitro} data of Section \ref{sec:invitro}, the optimization feasible region was chosen to be the same as the feasible region used in paper \cite{kohn2022phenotypic} for the \textit{Ba/F3} data, i.e.,
\begin{table}[ht]
    \centering
    \begin{tabular}{|c|c|c|c|c|c|c|c|c|}
    \hline
         & $p$ & $\beta$ & $\nu$ & $b$ & $E$ & $m$ & $\sigma_{L},\sigma_{H}$& c \\
         \hline
         Range & $[0,1]$ & $[0,1]$ & $[\beta-0.06,\beta]$ & $[0.878,1]$ & $[0,50]$& $[0.001,20]$ & $[0,2500]$& $[0,100]$. \\
         \hline
    \end{tabular}
    \caption{Optimization Feasible region}
    \label{table:OptimizationConstriantsInVitro}
\end{table}

We solved each optimization problem 500 times starting from randomly chosen initial points.

%%%%%%%%%%%%%%%%%%%% Proof of multi-type CLT %%%%%%%%%%%%%%%%%%%%%
\subsection{Proof of proposition \ref{prop:LiveImageCLT}}
\label{appx:LiveImageCLT Proof}

Given a set of time points $\mathcal{T} = \{t_1,\cdots,t_k\}$, we first show for any $t_j, 1\leq j\leq k$ that:
\[W_{n}(t_j) \Rightarrow Y(t_j) =  \sum_{\ell = 1}^j \sum_{g=1}^S \sqrt{p_g} e^{\lambda_g (t_j - t_{\ell})} e^{\lambda_g t_{\ell - 1}/2}V_g(t_{\ell} - t_{\ell -1}) \text{ as } n \to \infty\]
Where $V_g(t)$ is a random variable that has normal distribution $N(0,\sigma_g^2(t))$ and the $\sigma_g^2(t)$ here is the variance of subpopulation $g$ linear birth-death process defined in equation \eqref{eq:BD_Variance}.
Following an argument from Either and Kurtz \cite{ethier2009markov}, we have the decomposition
\begin{align*}
    W_{n}(t_j) &= \frac{1}{\sqrt{n}} \sum_{g = 1}^{S} \sum_{\ell = 1}^{j} e^{\lambda_g (t_j - t_{\ell})} X_g(t_{\ell}) - e^{\lambda_g (t_j - t_{\ell - 1})} X_g(t_{\ell - 1})\\
    &= \sum_{\ell = 1}^{j} \sum_{g = 1}^{S} \frac{1}{\sqrt{ n}} e^{\lambda_g(t_j - t_{\ell})} \left[X_g(t_{\ell}) - e^{\lambda_g(t_{\ell} - t_{\ell - 1})} X_g(t_{\ell - 1})\right]\\
    &= \sum_{\ell = 1}^{j} \sum_{g = 1}^{S} e^{\lambda_g(t_j - t_{\ell})} \left(\frac{p_g  n}{n} \right)^{1/2} \left(\frac{X_g(t_{\ell - 1})}{p_g  n} \right)^{1/2} X_g(t_{\ell - 1})^{-1/2} \sum_{m = 1}^{X_g(t_{\ell - 1})} \left[ B_g(t_{\ell} - t_{\ell - 1}) - e^{\lambda_g(t_{\ell} - t_{\ell - 1})} \right]\\
\end{align*}
By assuming the initial proportion $p_g$ for sub-type $g$ is independent of $n$ and using the Law of large numbers, as $n\rightarrow \infty$, we have
\[\left( \frac{X_g(t_{\ell - 1})}{p_g  n}  \right) \rightarrow \E[B_g(t_{\ell - 1})] = e^{\lambda_g t_{\ell - 1}} \quad a.s.\]
By assuming that the maximum number of time point $N_T$ and the length of time interval $t_i - t_j$ for any $i\geq j$ are both bounded and not depend on $n$, the Law of large numbers also assures that for any $\ell \in \{1,\cdots,N_T\}$ the $X_g(t_{\ell-1})$ will diverge to infinity when $n\rightarrow \infty$. Therefore, we may apply the Central Limit Theorem to the following term:
\[X_g(t_{\ell - 1})^{-1/2} \sum_{m = 1}^{X_g(t_{\ell - 1})} \left[  B_g(t_{\ell} - t_{\ell - 1}) - e^{\lambda_g (t_{\ell} - t_{\ell - 1})} \right] \Rightarrow V_g(t_{\ell} - t_{\ell - 1}) \sim N(0, \sigma_{g}^2(t_{\ell} - t_{\ell - 1})). \]
Thus, we conclude that
\[W_{n}(t_j) \Rightarrow \sum_{\ell = 1}^j \sum_{g =1}^S \sqrt{p_g} e^{\lambda_g (t_j - t_{\ell})} e^{\lambda_g t_{\ell - 1}/2} V_g(t_{\ell} - t_{\ell -1}) \text{ as } n \to \infty\]
Next we show that the random vector $\mathbf{W}$ converges to the random vector $\mathbf{Y}$, which has the multivariate normal distribution. We can obtain the distribution for $\mathbf{Y}$ from the independence between $V_i(t_{\ell} - t_{\ell - 1})$ and $V_j(t_{m} - t_{m - 1})$ for all $i,j \in \{1,\cdots,N_g\}, \ell,m\in \{1,\cdots, k\}$:
\[\mathbf{Y}= [Y(t_1),\cdots, Y(t_k)] \sim N(0,\Sigma)\]
where
\[\Sigma_{i,j} =  \sum_{\ell = 1}^{\min(i,j)} \sum_{g =1}^{S} p_g e^{\lambda_g (t_i - t_{\ell})} e^{\lambda_g (t_j - t_{\ell})} e^{\lambda_g t_{\ell - 1}} \sigma_{k}^2(t_{\ell} - t_{\ell - 1})\]
Then we use the Cramer Wold device, given a constant vector $a\in \mathbb{R}^k < \infty$, we have
\begin{align*}
    \langle a, \mathbf{W}\rangle &= \sum_{j = 1}^k a_j \sum_{\ell = 1}^{j} \sum_{g =1}^{S} \sqrt{p_g} e^{\lambda_g (t_j - t_{\ell})} \left(\frac{X_g(t_{\ell - 1})}{p_g  n} \right)^{1/2} X_g(t_{\ell - 1})^{-1/2} \sum_{m = 1}^{X_g(t_{\ell - 1})} \left[B_m(t_{\ell} - t_{\ell - 1}) - e^{\lambda_g (t_{\ell} - t_{\ell - 1})} \right]\\
    & \Rightarrow \sum_{j = 1}^{k} a_j \sum_{\ell = 1}^{j} \sum_{g =1}^{S} \sqrt{p_g} e^{\lambda_g (t_{j} - t_{\ell})} e^{\lambda_g t_{\ell - 1}/2} V_{i}(t_{\ell} - t_{\ell - 1})\\
    &= \sum_{j = 1}^{k} a_j Y(t_j) = \langle a , \mathbf{Y} \rangle
\end{align*}
Thus, we show that $\mathbf{W} \Rightarrow \mathbf{Y}$.

\subsection{Proof of proposition \ref{prop:LiveImageCLT} if initial proportions can go to 0 with $n$ }

\label{appx:LiveImageCLT Extension}
In proving Proposition \ref{prop:LiveImageCLT}, we made the assumption that the initial proportions $p_i$ for $i=1,\cdots,S$ are not dependent on the initial cell count $n$. In this sub-section, we aim to relax this assumption and demonstrate a similar result. Note we will assume that all other inputs are still independent of $n$, i.e., $S$ and $\mathcal{T}$.

In particular, we will allow $p_i$ to depend on $n$, and allow for $\limsup_{n}p_in<\infty.$ We denote two sets of subpopulations $F$ and $I$, where 
\begin{align*}
    F &= \{i \in \{1,\cdots,S\};\limsup_{n\to\infty}p_i n < \infty \}\\
    I &= \{i \in \{1,\cdots,S\};\lim_{n\to\infty}p_i n = \infty \}
\end{align*}
Due to $\sum_{i = 1}^S p_i = 1$, and $S$ being fixed with $n$, we know that the set $I$ must not be an empty set. Then following a similar pattern as the proof of the Proposition \ref{prop:LiveImageCLT}, we derive:

\begin{align*}
    W_{n}(t_j) &= \sum_{\ell = 1}^{j} \sum_{g = 1}^{S} e^{\lambda_g(t_j - t_{\ell})} \left(\frac{p_g  n}{ n} \right)^{1/2} \left(\frac{X_g(t_{\ell - 1})}{p_g  n} \right)^{1/2} X_g(t_{\ell - 1})^{-1/2} \sum_{m = 1}^{X_g(t_{\ell - 1})} \left[ B_g(t_{\ell} - t_{\ell - 1}) - e^{\lambda_g(t_{\ell} - t_{\ell - 1})} \right]\\
    &= \sum_{\ell = 1}^{j} \sum_{g \in I} e^{\lambda_g(t_j - t_{\ell})} \left(\frac{p_g  n}{ n} \right)^{1/2} \left(\frac{X_g(t_{\ell - 1})}{p_g  n} \right)^{1/2} X_g(t_{\ell - 1})^{-1/2} \sum_{m = 1}^{X_g(t_{\ell - 1})} \left[ B_g(t_{\ell} - t_{\ell - 1}) - e^{\lambda_g(t_{\ell} - t_{\ell - 1})} \right]\\
    & + \sum_{\ell = 1}^{j} \sum_{g \in F} e^{\lambda_g(t_j - t_{\ell})} \left(\frac{p_g  n}{ n} \right)^{1/2} \left(\frac{X_g(t_{\ell - 1})}{p_g  n} \right)^{1/2} X_g(t_{\ell - 1})^{-1/2} \sum_{m = 1}^{X_g(t_{\ell - 1})} \left[ B_g(t_{\ell} - t_{\ell - 1}) - e^{\lambda_g(t_{\ell} - t_{\ell - 1})} \right]\\
\end{align*}

Next, we may consider these two double sums separately. For $g\in I$, because the $p_g n$ diverges to infinity as $n\rightarrow \infty$, the first double sum will converge to the same limit we established in Proposition \ref{prop:LiveImageCLT}. For $g\in F$, $p_g n$ will stay bounded. Therefore, in the second double sum, we will have $\frac{p_g n}{n}$ converge to 0 as $n\rightarrow \infty$, which makes the second double sum vanish. In conclusion, we have
\[W_{n}(t_j) \Rightarrow \sum_{\ell = 1}^{j} \sum_{g\in I} \sqrt{p_g} e^{\lambda_g (t_j - t_{\ell})} e^{\lambda_g t_{\ell - 1}/2} V_g(t_{\ell} - t_{\ell -1}) \text{ as } n \to \infty.\]
Note that this convergence result will lead to the same realization in practice, i.e., we will define the covariance by summing over all subtypes. This is because in practice we only have $n<\infty$ and we cannot actually assume subpopulations have zero contribution to the covariance.

%%%%%%%%%%%%%%%%%%% End of the proof %%%%%%%%%%%%%%%%%%%%%%%%%%%%%
\subsection{Exact path likelihood computation}
\label{appx:Exact Path likelihood}
Here we show how to calculate the exact likelihood of a sample path observation of the total cell count of multiple heterogeneous birth-death processes. While we do not use this approach for likelihood evaluation in the current manuscript we report it here to show that it is not feasible.

We first consider the following joint probability of a homogeneous linear birth-death process:
$$
    \mathbb{P}(X(t_1) = x_1,\cdots, X(t_k) = x_k|X(t_0) = n, \theta_{BD}(2))  = \prod_{k = 1}^{N_t} \mathbb{P}(X(t_k) = x_k | X(t_{k-1}) = x_{k-1},\theta_{BD}(2)).
$$
For ease of notation define the transition probability $p_{i, j}(t_k - t_{k-1}) =\mathbb{P}(X(t_k) = j | X(t_{k-1}) = i,\theta_{BD}(2))$. It is important to note that evaluating $p_{i,j}(t_k - t_{k-1})$ is the most computation-demanding task when evaluating the joint probability. As a result, we mainly consider the number of evaluations of this transition probability. The analytical form for this transition probability was derived in \cite{bailey1991elements}:
\begin{align}
    \label{eq:BD_transition}
    p_{i,j}(t) = \sum_{k = 0}^{\min(i,j)} \binom{i}{k} \binom{i+j-k-1}{i-1} a(t)^{i-k} b(t)^{j -k} (1 - a(t) - b(t))^{k},
\end{align}
where 
\[a(t) = \frac{\nu(e^{(\beta-\nu)t} - 1)}{\beta e^{(\beta-\nu)t} -\nu}, b(t) = \frac{\beta(e^{(\beta - \nu)t} - 1)}{\beta e^{(\beta - \nu)t} - \nu}.\]
Note that numerical evaluation \eqref{eq:BD_transition} will be computationally expensive due to the presence of multiple factorial terms. We use a Gosper refined version of the Stirling formula \cite{gosper1978decision} to approximate these factorials
\[n! \approx \sqrt{\left(2n + \frac{1}{3}\right)\pi} n^n e^{-n}.\]
We find that this approximation leads to good performance in our examples. It is thus straightforward to evaluate the path likelihood for the case of a homogeneous linear birth-death process.

If we instead have observations of a sum of birth-death processes, the evaluation of the path likelihood is much more difficult. In particular, the sum of the birth-death processes is no longer a Markov process and we must therefore sum over possible values of our unobserved subpopulations.
Specifically, if we have two subpopulations we can formulate the equation \eqref{eq:LC path likelihood}, as 
\begin{equation}
\label{eq:naive_transition}
\begin{split}
&P(X^{(r)}(t)=x_t, t\in\mathcal{T}|\theta_{BD}(2))\\
&=\sum_{i_1=0}^{x_1}\cdots\sum_{i_{N_T}=0}^{x_{N_T}}P\left(X_1(t_1)=i_1,X_2(t_1)=x_1-i_1,\ldots,X_1(t_{N_T})=i_{N_T},X_2(t_{N_T})=x_{N_T}-i_{N_T}|\theta_{BD}(2)\right)\\
&=
\sum_{i_1=0}^{x_1}\cdots\sum_{i_{N_T}=0}^{x_{N_T}}P\left(X_1(t_1)=i_1,X_1(t_2)=i_2,\cdots,X_1(t_{N_T}) = i_{N_T}|\theta_{BD}(2)\right)\\
&\quad\times P\left(X_2(t_1)=x_1 - i_1,X_2(t_2)=x_2-i_2,\cdots, X_2(t_{N_T}) = x_{N_{T}} - i_{N_T}|\theta_{BD}(2)\right).
\end{split}
\end{equation}
Note that the last equality is due to the assumption that subpopulation grow independently, and we can decompose the joint probability of mixture cell count into a summation of multiple joint probabilities of the homogeneous linear birth-death process. It is not hard to see that if we naively evaluate the above sum, the number of computations of the homogeneous joint probability is around $\Omega(\min_{t\in\tau} x_t^{N_T})$.
Since many examples have $N_T\approx 10$, and $x_t\approx 100$ this is clearly an infeasible approach.

% \subsection{Hidden Markov Model}
% \label{HMM}

%%%%%%%%%%%%%%%%%%%%%%%%%%% HMM send to appendix %%%%%%%%%%%%%%%%%%%%%%%%%%%%%%%
In order to avoid the exponential dependence on the number of time points, one option is to use techniques from Hidden Markov Models (HMM). The main assumption of HMM is the Markov property of the hidden process, and that the hidden process relates to the observable process according to a specified distribution $B$. Recall that the time series of observed total cell count is given by $\{X(t_i);i\in\mathcal{T}\}$ and denote the time series of the subpopulations as $\{(X_1(t_i),\ldots,X_S(t_i);i\in\mathcal{T}\}$. Then $\{X(t_i);i\in\mathcal{T}\}$ in the HMM is the observable process, and $\{(X_1(t_i),\ldots,X_S(t_i));i\in\mathcal{T}\}$ is the hidden Markov process due to the Markov property of the linear birth-death process. Notice that the relationship between the hidden process $\{(X_1(t_i),\ldots,X_S(t_i));i\in\mathcal{T}\}$ and the observable process $\{X(t_i);i\in\mathcal{T}\}$ can be defined as \[\mathbb{P}(X(t) = x | (X_1(t),\ldots,X_S(t)) = (x_1,\ldots,x_S)) = \begin{cases}
1 & \text{ if } x_1 +\ldots+ x_S = x\\
0 & \text{ o.w. }
\end{cases}\]

We can translate the live-cell imaging experiment into an HMM and significantly improve the computational complexity of evaluating the exact likelihood function \eqref{eq:BD_LikelihoodFun}. In particular, we can use popular HMM techniques, such as the forward-backward procedure to reduce the total number of transition probability, i.e., equation \eqref{eq:BD_transition}, computation to $\Theta(H^2 N_T)$ for one replicate at one dosage level, where $H$ is the number of hidden states. In particular, we need to calculate the $H$ by $H$ transition matrix for every time point, and if we assume the length of time intervals are identical, we can reduce the upper bound to $\Theta(H^2)$. However, the number of hidden states depends on both the maximum total number of cells observed at each time point $x$ and the number of subpopulations $S$. As we will now show, this is unfortunately not a sufficient reduction in computational complexity. In particular, assume we observe $x$ total cells. In that case, the number of hidden states is given by $\binom{x-1}{S-1}$, and assuming that $x\gg S$, we have that $\binom{x-1}{S-1}=\Theta\left(x^{S-1}\right)$ as $x\to\infty$. With only two subpopulations this results in computational complexity of $\Theta(x^2)$, and in many experiments, we might have $x\approx 10^5$ leading to an extremely high computational burden. If $S=3$ we would end up with the far worse computational complexity of $\Theta(x^4)$.

In conclusion, using a naive approach to compute the joint probability of mixture cell count would require $\Omega\left(\min_{t\in\mathcal{T}}x_t^{N_T}\right)$ many computations of transition probability, which is clearly infeasible when many cell counts are in the thousands with over 10 observed time points. We also discuss an HMM based approach to evaluating the likelihood that results in a significant reduction in the computational burden for evaluating this likelihood. In particular, with this approach we can reduce the number of computations of the transition probability to $\Omega\left(\min_{t\in\mathcal{T}}x_t^2\right)$. Unfortunately we might have $\min_{t\in\mathcal{T}}x_t\approx 5000$. Note that this will be the complexity for evaluating the likelihood of one single replicate at a single dose, so taking into account that we can have more than 10 different doses, with more than 10 replicates at each dose we see that unfortunately, this HMM approach is computationally infeasible. In addition, this HMM approach will have significantly worse computational complexity after we include observation noise terms and/or more than 2 subpopulations.

\subsection{Data and code availability}
% In Figures \ref{fig:CI width}, \ref{fig:CI width 10 noise}, and \ref{fig:CI width 50 noise}, statistics were done using R version 4.2.1 \cite{R4.2.1} with ggsignif package \cite{ggsignif}.
All data and code used for running experiments, model fitting, and plotting is available on a GitHub repository at \url{https://github.com/chenyuwu233/PhenoPop_stochastic}. The required Matlab version is Matlab R2022a or newer.

\section*{Acknowledgements}
The work of C. Wu was supported in part  with funds from the Norwegian Centennial Chair Program. The work of K. Leder was supported in part with funds from NSF award CMMI 2228034 and Research Council of Norway Grant 309273. The work of EBG and JF was supported in part by NIH grant R01 CA241137, NSF DMS 2052465, NSF CMMI 2228034, and Research Council of Norway Grant 309273. The work of J.M. Enserink and D.S. Tadele was supported by grants from the Norwegian Health Authority South-East, grant numbers 2017064, 2018012, and 2019096; the Norwegian Cancer Society, grant numbers 182524 and 208012; and the Research Council of Norway through its Centers of Excellence funding scheme (262652) and through grants and 261936, 294916 and 314811

    \bibliographystyle{plain}
    \bibliography{main.bib}

\end{document}